\documentclass[%
 reprint,
 amsmath,amssymb,
 aps,
]{revtex4-1}
\pdfoutput=1
\usepackage{graphicx}
\usepackage{amsmath}
\usepackage{amssymb}
\usepackage{dsfont}
\usepackage{tensor}
\usepackage{upgreek}
\usepackage{braket}
\usepackage{hyperref}
\usepackage[letterpaper,left=1.5cm,right=1.5cm,top=1.6 cm,bottom=2.5cm]{geometry}

\begin{document}
\pdfoutput=1
\preprint{APS/123-QED}

\title{Spin-lattice relaxation of individual solid-state spins}

\author{{ A. Norambuena$^{1,2}$, E. Mu\~{n}oz$^{1,2}$, H. T. Dinani$^1$, A. Jarmola$^{3}$, P. Maletinsky$^{4}$, D. Budker$^{3,5,6}$, and J. R. Maze$^{1,2}$}\\
\vspace{0.5 cm}
{\small \em $^1$ Faculty of Physics, Pontificia Universidad Cat\'olica de Chile, Avda. Vicu\~{n}a Mackenna 4860, Santiago, Chile\\
$^2$ Center for Nanotecnology and Advanced Materials CIEN-UC, Pontificia Universidad Cat\'olica de Chile, Avda. Vicu\~{n}a Mackenna 4860, Santiago, Chile \\
$^3$ Department of Physics, University of California, Berkeley 94720-7300, USA \\
$^4$ Department of Physics, University of Basel, Klingelbergstrasse 82, CH-4056 Basel, Switzerland  \\
$^5$ Helmholtz Institute, Johannes Gutenberg University, 55128 Mainz \\
$^6$ Nuclear Science Division, Lawrence Berkeley National Laboratory 
}}
\date{\today}% It is always \today, today,

\begin{abstract}
Understanding the effect of vibrations on the relaxation process of individual spins is crucial for
implementing nano systems for quantum information and quantum metrology applications. In this
work, we present a theoretical microscopic model to describe the spin-lattice relaxation of individual
electronic spins associated to negatively charged nitrogen-vacancy centers in diamond, although our
results can be extended to other spin-boson systems. Starting from a general spin-lattice interaction
Hamiltonian, we provide a detailed description and solution of the quantum master equation of an
electronic spin-one system coupled to a phononic bath in thermal equilibrium. Special attention
is given to the dynamics of one-phonon processes below 1 K where our results agree with recent
experimental findings and analytically describe the temperature and magnetic-field scaling. At
higher temperatures, linear and second-order terms in the interaction Hamiltonian are considered
and the temperature scaling is discussed for acoustic and quasi-localized phonons when appropriate.
Our results, in addition to confirming a $T^5$ temperature dependence of the longitudinal relaxation
rate at higher temperatures, in agreement with experimental observations, provide a theoretical
background for modeling the spin-lattice relaxation at a wide range of temperatures where different
temperature scalings might be expected.

\begin{description}
%\item[References]	
\item[PACS numbers]
\end{description}
\end{abstract}

\pacs{Valid PACS appear here}% PACS, the Physics and Astronomy
\maketitle

%\tableofcontents

\section{Introduction}
The negatively charged nitrogen-vacancy (NV$^{-}$) center in diamond is a promising solid-state system with remarkable
applications in quantum sensing with atomic-scale spatial resolution \cite{MazeNature2008,Ajoy2015}, fluorescent marking of biological structures \cite{Fu2007, Faklaris2009,McGuinness2011}, single photon sources \cite{Naydenov2014}, and quantum communications \cite{Fuchs2006}. However, most of these quantum-based applications crucially depend on the longitudinal ($1/T_1$) and transverse ($1/T_2$) spin relaxation rates associated with the
ground state spin degree of freedom \cite{Jarmola2012}. \par

From experiments and theory, we know that lattice phonons in diamond are important for the spin-lattice relaxation
dynamics of the spin degree of freedom of the NV$^{-}$ center, and that the temperature plays a fundamental role
in this relaxation process \cite{Redman1991,Harrison2006,Takahashi2008,Jarmola2012,Astner2017}. Phonons can be understood
as collective quantum vibrational excitations that propagate through the lattice and directly interact with
the orbital states of the point defect. The intensity of this interaction depends on the electron-phonon coupling between
the defect and all possible phonon modes in the lattice (acoustic, optical and quasi-localized phonon modes) \cite{Alkauskas2014,Londero2016,Ariel2016}. Theoretical and numerical studies show that the strain field of the diamond lattice and perturbative corrections given by the spin-orbit and spin-spin interactions introduce interesting spin-phonon dynamics between the ground state spin degree of freedom of the NV$^{-}$ center and lattice phonons \cite{Doherty2011, Doherty2013}.\par

Several theoretical works have addressed the problem of finding the relaxation rate by considering the interaction between the spin degree of freedom with two-phonon Raman \cite{Vleck1940,Walker1968} and Orbach-type \cite{Abragam1970} processes. In general, the problem of estimating the thermal dependence of each relaxation process is translated into the problem of calculating the transition rates predicted by the Fermi golden rule for different phonon processes \cite{Vleck1940,Walker1968,Abragam1970,Astner2017}. Using this reasoning, it is reported that the second-order Raman process induced by a linear spin-phonon interaction leads to $1/T_1 \propto T^{5}$ \cite{Walker1968}, while the first-order Raman process induced by a quadratic spin-phonon interaction leads to $1/T_1 \propto T^{7}$ \cite{Vleck1940}, where $T$ is the environment temperature. \par

The ground triplet state of the NV$^{-}$ center in diamond has a natural zero-field splitting $D/2\pi = 2.87$ GHz originated from the dipole-dipole interaction between electronic spins \cite{Ivady2014, Doherty22014}. This energy gap is low compared to typical optical phonon energies $\omega_{\mbox{\scriptsize ph}}/2\pi \sim$ 15-40 THz and sets a characteristic thermal gap associated with the spin system $T_{\mbox{\scriptsize gap}} = \hbar D/k_B \approx 0.14$ K. Experimental observations at high temperatures, from 300 K to 475 K, have shown that different samples with different NV$^{-}$ center concentrations present a dominant two-phonon Raman process that leads to $\left(1/T_1\right)_{\mbox{\scriptsize Raman}} \propto T^5$ \cite{Takahashi2008,Jarmola2012}. At low temperatures, between 4 K and 100 K, the relaxation rate is dominated by Orbach and spin-bath interactions. The former is associated with a quasi-localized phonon mode with energy $\omega_{\mbox{\scriptsize loc}} \approx$ 73 meV \cite{Jarmola2012,Huxter2013} and contributes with a temperature dependence relaxation rate $(1/T_1)_{\mbox{\scriptsize Orbach}} \propto (\mbox{exp}(\hbar \omega_{\mbox{\scriptsize loc}} /k_B T)-1)^{-1}$. This, closely matches the numerical vibrational resonance predicted by \textit{ab initio} studies \cite{Alkauskas2014}. Meanwhile, it is observed that dipole-dipole interactions between neighboring spins lead to a constant sample-dependent relaxation rate which dominates at this temperature range \cite{Jarmola2012}. In contrast, at lower temperatures (below 1 K) recent experimental observations and \textit{ab initio} calculations concluded that the longitudinal relaxation rate is dominated by single-phonon processes, and is given
by $(1/T_1) \propto \Gamma_0\left(1+3\bar{n}(T)\right)$, where $\Gamma_0 = 3.14 \times 10^{-5}$ s$^{-1}$, and $\bar{n}(T) = \left(\mbox{exp}(\hbar D / k_B T)-1 \right)^{-1}$ is the mean number of phonons at the zero-field splitting frequency \cite{Astner2017}. However, a microscopic model that predicts the temperature dependence of the longitudinal relaxation rate for a wide range of temperatures, to the best of our knowledge, is still missing. \par

Here, we present a microscopic model for the spin-lattice relaxation dynamics associated with the ground state of
the NV$^{-}$ center in diamond. In our model, we introduce a general spin-phonon Hamiltonian to describe the spin relaxation
dynamics using the quantum master equation associated with the electronic spin degree of freedom under
the effect of a phononic bath. We focus on the estimation of the longitudinal relaxation rate by evaluating the rate
of the Fermi golden rule transitions to first and second-order considering the effect of acoustic and quasi-localized
phonons. In Sec. II, we give the Hamiltonian of the whole system and introduce the spin-phonon interaction between
the triplet state of the spin degree of freedom and lattice vibrations, by considering one-phonon and two-phonon interactions.
Section III introduces the phonon damping rates for one-phonon and two-phonon processes, by using the
Fermi golden rule, the Debye approximation, and a model for strong interactions with quasi-localized phonon modes.
In Sec. IV we introduce the quantum master equation associated with the spin-lattice relaxation dynamics of the ground
state and include the role of a stochastic magnetic noise. Finally, in Section V we discuss the longitudinal relaxation rate at
low and high-temperature regimes and the role of a static magnetic field on the relaxation rate for low temperatures.

\section{Spin degree of freedom and phonons}
We consider a system composed of a single NV$^{-}$ center in diamond interacting with lattice phonons. In this
scenario, local vibrations induce a mixing between orbital states of the defect by means of the electron-phonon interaction.
This phonon-induced mixing effect generates an effective interaction between the spin degree of freedom and
lattice phonons. In order to model the spin-phonon relaxation dynamics, we use the following Hamiltonian
\begin{equation}
\hat{H} = \hat{H}_{\mbox{\scriptsize NV}} + \hat{H}_{\mbox{\scriptsize s-ph}} + \hat{H}_{\mbox{\scriptsize ph}},
\end{equation}
where the first, second and third terms represent the ground state spin Hamiltonian of the NV$^{-}$ center, the interaction
Hamiltonian between the spin state and lattice phonons, and the phonon bath, respectively. \par

The NV$^{-}$ center is composed of a substitutional nitrogen atom next to a vacancy in a diamond lattice. The symmetry
of the center is captured by including the three carbon atoms adjacent to the vacancy \cite{Gali2008}. The atomic configuration
of this point defect is associated with the $C_{3v}$ symmetry group. The electronic structure of this point defect is modeled as a two electron-hole system with electronic spin $S = 1$. In this representation, the electronic wavefunctions
of the excited and ground state are linear combinations of two-electron wave functions \cite{Larsson2008}, where the single-electron orbitals of the NV$^{-}$ center can be written in terms of the carbon and nitrogen dangling bonds \cite{Loubser1978,Goss1996}. In the absence of external perturbations, such as lattice distortions
or electromagnetic fields, the orbital excited states $\ket{X}$ and $\ket{Y}$ are degenerate due to the $C_{3v}$ symmetry and belong to the irreducible representation $E$. Meanwhile, the orbital ground state $\ket{A_2}$ belongs to the irreducible representation $A_2$. \par

\begin{figure}
\includegraphics[width= 1 \linewidth]{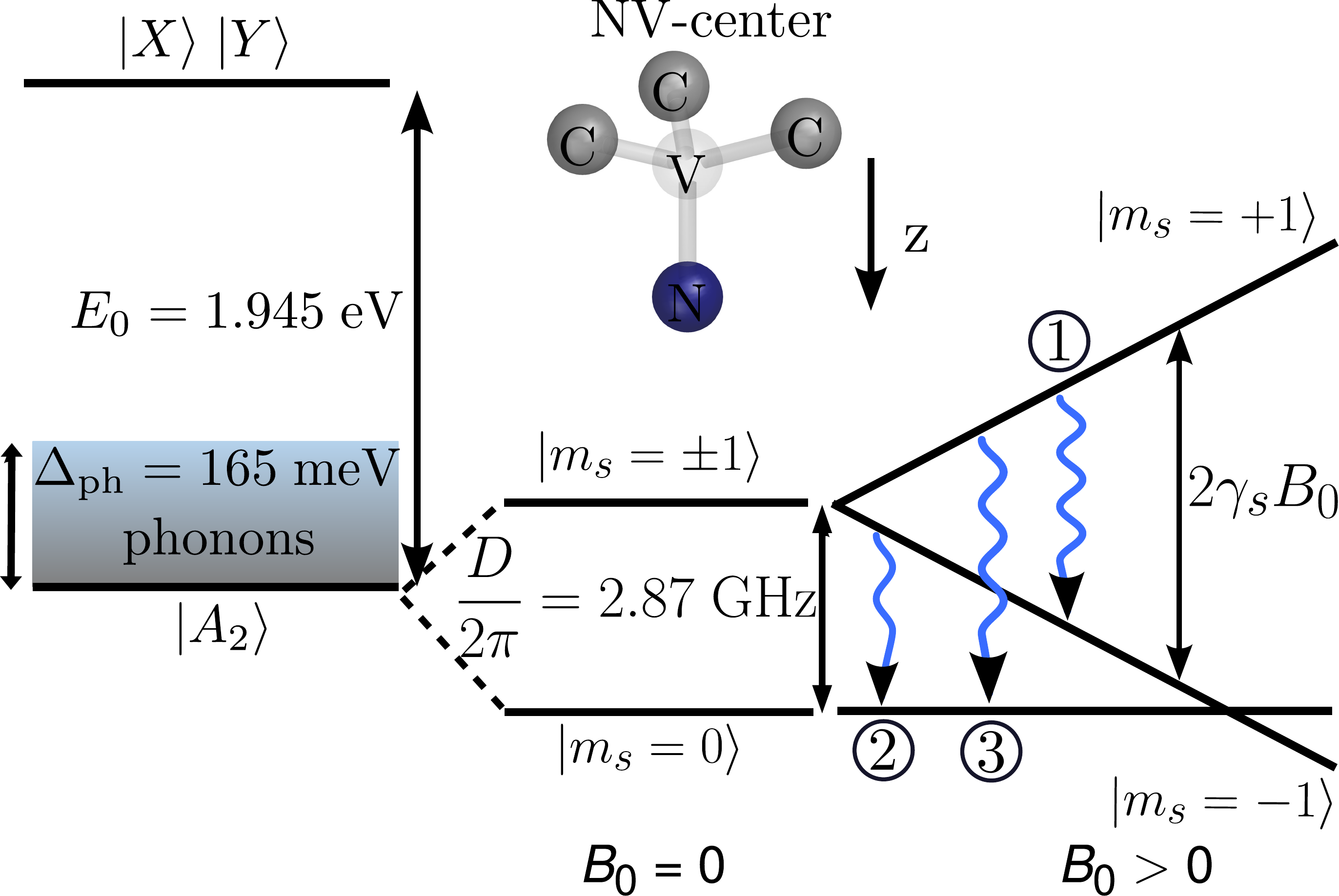}
\caption{The energy levels and the atomic structure of the NV$^{-}$ center are shown. Here, $\ket{X}$ and 
$\ket{Y}$ are the orbital degenerate excited states, and $\ket{A_2}$ is the orbital ground state. The zero-phonon line energy
is given by $E_0 = 1.945$ eV. The spin triplet states are represented by $\ket{m_s = 0}$ and $\ket{m_s = \pm 1}$. Such spin states are separated by the zero-field splitting constant $D/2\pi =  2.87$ GHz and the static magnetic field which we have assumed is aligned along the symmetry axis of the center. Phonons are represented by a continuous band that interacts with the ground state and its transitions are represented by the labels (1), (2), and (3).}
\label{fig:Figure1}
\end{figure}

In the presence of a static magnetic field $B_0$ along the $z$ axis, the spin Hamiltonian of the NV$^{-}$ center is given by ($\hbar = 1$)
\begin{equation}
\hat{H}_{\mbox{\scriptsize NV}} =  D S_z^2 + \gamma_s B_0 S_z, \label{SpinHamiltonian}
\end{equation}
where $\mathbf{S} = (S_x,S_y,S_z)$ are the Pauli matrices for $S = 1$ (dimensionless), $D/2\pi = 2.87$ GHz is the zero-field splitting constant, and $\gamma_s/2\pi \approx 2.8$ MHz/G is the gyromagnetic ratio. Figure.~\ref{fig:Figure1} shows the energy
diagram of the system, including the orbital states, spin degrees of freedom and the atomic configuration of the NV$^{-}$
center. \par

Quantum systems with spin $S = 1$ are traditionally called non-Kramers systems \cite{Mueller1968, CharlesPoole1970}.
Interestingly, there is a non-trivial connection between the spin number and the temperature dependence of the relaxation
rate \cite{Walker1968,Abragam1970,Loubser1978}. Therefore, in order to obtain the correct temperature dependence of the spin relaxation rate of the ground triplet state of the NV$^-$ center we consider the most general spin-phonon interaction Hamiltonian for spin $S=1$ systems given by \cite{Mueller1968}
\begin{eqnarray}
 \hat{H}_{\mbox{\scriptsize s-ph}} &=&  E_z S_z^2 + E_x \left(S_x^2 -S_y^2 \right) + E_{y}\left(S_x S_y + S_y S_x \right) \nonumber \\
&& + E_{x'}\left(S_x S_z + S_z S_x \right)   + E_{y'}\left(S_y S_z + S_z S_y \right), \label{Spin-Phonon}
\end{eqnarray}
where the operators $E_{z}, E_{x},E_{y},E_{x'}$ and $E_{y'}$ have units of energy. In addition, the operators $E_{x}$, $E_{x'}$, 
$E_{y}$, and $E_{y'}$ belong to the irreducible representation $E$, while the term $E_z$ is characterized by the irreducible representation $A_1$ \cite{Mueller1968}. Physically, the $E_i$ operators can be derived from perturbative corrections of the 
spin-spin and spin-orbit interactions due to the effect of the strain field \cite{Doherty2013}. These operators are proportional to the nuclear displacements, and therefore, can be quantized using phonon modes \cite{Doherty2013}. In order to introduce these quantized vibrations, we expand the $E_i$ operators in terms of symmetric lattice phonon-mode operators, including the linear and the quadratic terms, as the following
\begin{eqnarray}
E_i &=& \sum_{k \in E} \lambda_{k,i} \hat{x}_k+ \sum_{k \otimes k' \in E} \lambda_{k k',i} \hat{x}_k \hat{x}_{k'},  \quad i \neq z \label{PhononExpansionE} \\
E_z &=& \sum_{k \in A_1} \lambda_{k,z} \hat{x}_k + \sum_{k \otimes k' \in A_1} \lambda_{k k',z} \hat{x}_k \hat{x}_{k'}.  \label{PhononExpansionA1}
\end{eqnarray}
Here, $\lambda_{k,i}$ and $\lambda_{kk',i}$ are the linear and quadratic spin-phonon coupling constants, respectively. The operator $\hat{x}_k$ is given by $\hat{x}_k = \hat{b}_k+\hat{b}^{\dagger}_k$ where $\hat{b}_k$ and $\hat{b}^{\dagger}_k$ are the boson annihilation and creation operators, respectively satisfying $[\hat{b}_k,\hat{b}^{\dagger}_{k'}] = \delta_{k,k'}$. The linear term given in Eqs.~\eqref{PhononExpansionE} and \eqref{PhononExpansionA1} has the same symmetry as the corresponding $E_i$ operators, and phonons with these symmetry are considered in the summation. In the quadratic term we are considering combinations of phonons such that the product belongs to the irreducible representation $E$ or $A_1$. As a consequence of the multiplication rules $A_2 \otimes A_2 = A_1$ and $A_2 \otimes E = E$, phonon modes with $A_2$ symmetry only contribute to the quadratic term. Therefore, the most general spin-phonon Hamiltonian for a system with spin $S=1$, is given by
\begin{eqnarray}
 \hat{H}_{\mbox{\scriptsize s-ph}} &=& \sum_{i} \left[\sum_{k \in \Gamma_i }\lambda_{k,i} \hat{x}_k + 
 \sum_{k \otimes k' \in \Gamma_i }\lambda_{k k',i} \hat{x}_k \hat{x}_{k'} \right] \hat{F}_i(\mathbf{S}), \nonumber \\
 \label{Final-Spin-Phonon-Hamiltonian}
\end{eqnarray}
where $i = x,y,x',y',z$ is the spin label, $\Gamma_{x,y,x',y'} = E$ and $\Gamma_z = A_1$ are the irreducible representations of the $C_{3v}$ point group. The spin functions are given by $\hat{F}_{x}(\mathbf{S}) = S_x^2 -S_y^2, \hat{F}_{y}(\mathbf{S}) = S_x S_y + S_y S_x , \hat{F}_{x'}(\mathbf{S}) = S_x S_z + S_z S_x, \hat{F}_{y'}(\mathbf{S}) = S_y S_z + S_z S_y$, and $\hat{F}_{z}(\mathbf{S}) = S_z^2$. \par

Using the spin basis that diagonalize the spin Hamiltonian given in Eq.~\eqref{SpinHamiltonian}, i.e., $\ket{m_s =1} = (1, 0, 0), \ket{m_s =0} = (0, 1, 0)$, and $\ket{m_s =-1} = (0, 0, 1)$ we explicitly obtain
\begin{eqnarray}
\hat{F}_x(\mathbf{S}) &=& \left(\begin{array}{rrr}
                                                                 0 &  0 &1  \\
                                                                0 &  0 & 0  \\
                                                                1 &0  &  0
                                                           \end{array} \right), \; 
\hat{F}_{x'}(\mathbf{S}) = {1 \over \sqrt{2}}\left(\begin{array}{rrr}
                                                                 0 &  1 & 0  \\
                                                                1 &  0 & -1  \\
                                                                 0 &-1  &  0
                                                           \end{array} \right), \\                                                           
 \hat{F}_y(\mathbf{S}) &=& \left(\begin{array}{rrr}
                                                                 0 & 0 & -i \\
                                                                0  &   0 &  0\\
                                                               i   & 0   & 0
                                                           \end{array} \right),  \; 
 \hat{F}_{y'}(\mathbf{S}) = {1 \over \sqrt{2}}\left(\begin{array}{rrr}
                                                                 0 & -i & 0 \\
                                                                i &   0 & i\\
                                                               0 & -i   & 0
                                                           \end{array} \right), \\
\hat{F}_z(\mathbf{S}) &=& \left(\begin{array}{rrr}
                                                                 1 & 0 & 0 \\
                                                                0  &   0 &  0\\
                                                              0 & 0   & 1
                                                           \end{array} \right).                                                            
\end{eqnarray}  
We observe that only the terms $\hat{F}_x(\mathbf{S})$ and $\hat{F}_y(\mathbf{S})$ induce spin transitions between the states 
$m_s = +1$ and $m_s = -1$, where the selection rule is $\Delta m_s = \pm 2$. On the other hand, the terms $\hat{F}_{x'}(\mathbf{S})$ and $\hat{F}_{y'}(\mathbf{S})$ induce spin transitions between $m_s = \pm 1$ and $m_s = 0$, in this case the selection rule is $\Delta m_s = \pm 1$. \par

Finally, the phonon Hamiltonian can be written as
\begin{eqnarray}
\hat{H}_{\mbox{\scriptsize ph}} &=& \sum_{k}\hbar \omega_ k \hat{b}_{k}^{\dagger} \hat{b}_k, \label{PhononHamiltonian}
\end{eqnarray}
where $\omega_k$ is the frequency of each vibrational mode of the lattice (including the color center), and the summation
takes into account the contribution of all phonon modes of the diamond lattice. In the next section, we will introduce
the phonon-induced spin relaxation rates and the temperature dependence associated to the spin-phonon Hamiltonian given
in Eq.~\eqref{Final-Spin-Phonon-Hamiltonian} by considering the effect of acoustic and quasi-localized
phonons in thermal equilibrium. We will show that the dimension and the symmetry of the lattice plays
a fundamental role in the temperature dependence of the longitudinal relaxation rate for two-phonon processes.

\section{Fermi Golden rule and phonon-induced spin relaxation rates}
In order to formally introduce the phonon-induced relaxation rates, we use the Fermi golden rule to first and second order by
using the spin-phonon Hamiltonian given in Eq.~\eqref{Final-Spin-Phonon-Hamiltonian}. To model first and second-order Raman-like processes, as well as direct absorption and emission associated with one-phonon processes. In particular, the energies associated with the spin transitions in the
ground state of the NV$^{-}$ center are given by $\omega_1 = 2\gamma_s B_0$, $\omega_2 = D + \gamma_s B_0$, and $\omega_3 = D - \gamma_s B_0$. For typical magnitudes of the static magnetic field $B_0 \sim 0-2000 \; \mbox{G}$ and taking into account
the zero field splitting constant $D/2\pi = 2.87$ GHz, we obtain that $\omega_1 \sim 0-11.2 \; \mbox{GHz}$, $\omega_{2,3} \sim 2.87-8.47 \; \mbox{GHz}$. These are the typical energies of acoustic phonons which belong to the linear branch of the phonon dispersion relation for diamond \cite{Pavone1993}. Acoustic phonons in diamond has energies of the order of $\omega_{\mbox{\scriptsize acous}} \sim 0-10$ THz. Therefore, the main fraction of acoustic phonons satisfy the frequency condition $\omega_{\mbox{\scriptsize acous}} \gg \omega_i$. \par

For the case of Raman-like processes the frequency condition is $\omega_{\mbox{\scriptsize ph,1}} - \omega_{\mbox{\scriptsize ph,2}} = \omega_{i}$ ($i=1,2,3$). Due to the condition $\omega_{\mbox{\scriptsize acous}} \gg \omega_i$ we assume in our model that the most significant contribution to two-phonon processes comes from acoustic phonons that satisfy $\omega_{\mbox{\scriptsize ph,1}}\gg \omega_i$ and $\omega_{\mbox{\scriptsize ph,2}}\gg \omega_i$. On the other hand, high energy phonons in diamond, with frequencies of the order of $\omega_{\mbox{\scriptsize ph}} \sim 15-40$ THz, can be included by considering the strong interaction with quasi-localized phonons. Therefore, in what follows we will consider the contribution of acoustic and quasi-localized phonons. \par

\subsection{One-phonon processes: acoustic phonons}
In the case of one-phonon processes, we need to distinguish between the absorption and the emission of a particular phonon mode with frequency $\omega_k$, which must be resonant with a transition between the spin energy levels of the NV$^{-}$ center in diamond. In order to introduce the temperature, we assume a phonon environment in thermal equilibrium, i.e., phonons that satisfy the Bose-Einstein distribution. Thus, we have $\langle \hat{b}_{k}^{\dagger} \hat{b}_k \rangle = n(\omega_k)$ and $\langle \hat{b}_{k} \hat{b}_k^{\dagger} \rangle = 1+ n(\omega_k)$, where $n(\omega_k) = \left[\mbox{exp}\left(\hbar \omega_k / k_B T \right) - 1 \right]^{-1}$ is the mean number of phonons at thermal equilibrium with $k_B$ and $\hbar$ being the Boltzmann and Planck constant, respectively. \par

\begin{figure}
\includegraphics[width= 1 \linewidth]{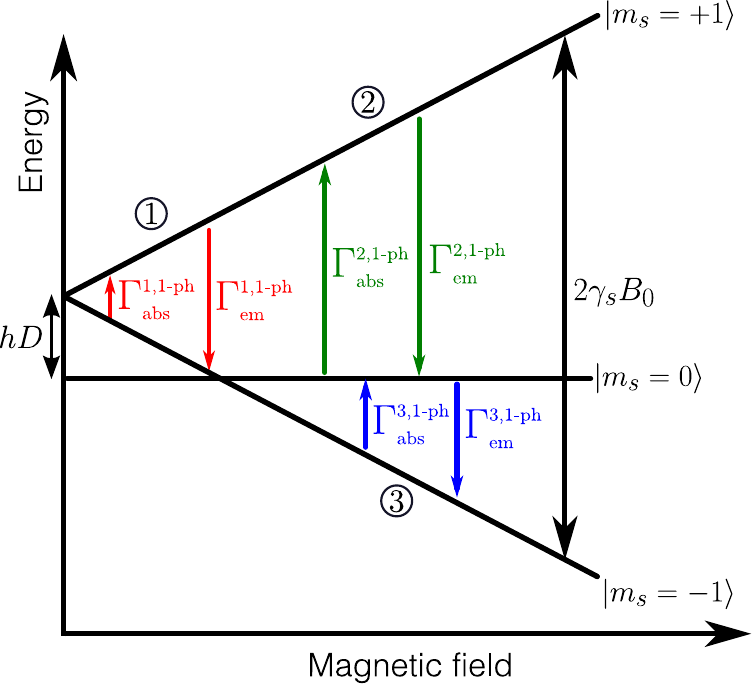}
\caption{The solid black lines are the energy levels of the ground triplet state of the NV$^{-}$ center in diamond as a function of the external magnetic field along the $z$ axis. For a given absorption and emission transition between two spin states $|m_s\rangle$, we observe three different spin relaxation processes represented by colored arrows (1=red, 2=green and 3=blue). The one-phonon damping rates $\Gamma^{i,\mbox{\tiny 1-ph}}_{\mbox{\tiny abs}}$ and $\Gamma^{i,\mbox{\tiny 1-ph}}_{\mbox{\tiny em}}$ are the absorption and emission spin relaxation rates for one-phonon processes.}
\label{fig:Figure2}
\end{figure}

For one-phonon processes the absorption and emission transition rates associated with the spin transition $\ket{m_s} \rightarrow \ket{m_s'}$ are given by the first order Fermi golden rule as
\begin{eqnarray}
\Gamma^{m_s \rightarrow m_s'}_{\mbox{\scriptsize abs}} &=& {2\pi \over \hbar^2}\sum_{k} \left| \langle m_s',n_k-1 \left| \hat{H}_{\mbox{\scriptsize s-ph}} , \right| m_s,n_k \rangle\right|^2 \nonumber \\ 
&& \hspace{1 cm} \times \delta(\omega_{m_s',m_s} -\omega_k), \\
\Gamma^{m_s \rightarrow m_s'}_{\mbox{\scriptsize em}} &=& {2\pi \over \hbar^2}\sum_{k} \left| \langle m_s',n_k+1 \left| \hat{H}_{\mbox{\scriptsize s-ph}}  \right| m_s,n_k \rangle\right|^2 \nonumber \\ 
&& \hspace{1 cm} \times \delta(\omega_{m_s',m_s} -\omega_k),
\end{eqnarray}
where $\omega_{m_s',m_s} = \omega_{m_s'}-\omega_{m_s}$ is the frequency difference between the spin sub-levels, and 
$\ket{n_k}$ is the number of phonons in the mode $k$ (Fock state). Using the spin-phonon Hamiltonian given in Eq.~\eqref{Final-Spin-Phonon-Hamiltonian}, the spin relaxation rates associated with one-phonon processes are given by
\begin{flalign}
&\Gamma_{\mbox{\scriptsize abs}}^{1, \mbox{\scriptsize 1-ph}} = {2\pi \over \hbar^2}n(\omega_1)J_{1}(\omega_1), \hspace{0.2 cm}
\Gamma_{\mbox{\scriptsize em}}^{1, \mbox{\scriptsize 1-ph}} = {2\pi \over \hbar^2}\left(n(\omega_1)+1\right)J_{1}(\omega_1),& \\
&\Gamma_{\mbox{\scriptsize abs}}^{2, \mbox{\scriptsize 1-ph}} ={\pi \over \hbar^2}n(\omega_2)J_{2}(\omega_2), \hspace{0.2 cm}
\Gamma_{\mbox{\scriptsize em}}^{2, \mbox{\scriptsize 1-ph}}  = {\pi \over \hbar^2}\left(n(\omega_2)+1\right)J_{2}(\omega_2),& 
\end{flalign}
where the subscripts $``1"$ and $``2"$ represent the spin transitions $\ket{m_s = -1} \leftrightarrow \ket{m_s = 1}$ and 
$\ket{m_s = 0} \leftrightarrow \ket{m_s = +1}$, respectively. Here, $J_1(\omega)$ and $J_2(\omega)$ are the spectral density functions 
\begin{eqnarray}
J_{1}(\omega) &=&  \sum_{k \in E}\left(\lambda_{k,x}^2+ \lambda_{k,y}^2\right) \delta(\omega-\omega_k), \\
J_{2}(\omega) &=&  \sum_{k \in E}\left(\lambda_{k,x'}^2+\lambda_{k,y'}^2\right) \delta(\omega-\omega_k), 
\end{eqnarray}
where $\lambda_{k,i}$ are the linear spin-phonon coupling constants, $\omega_k$ are the phonon frequencies, and both summations consider the contribution of $E$ phonons. For the transition $\ket{m_s = 0} \leftrightarrow \ket{m_s = -1}$ the gap frequency 
$\omega_3 =  D - \gamma_s B_0$ can be positive or negative depending on the strength of the external magnetic field $B_0$. For 
$\omega_3 > 0$ the absorption and emission damping rates are given by
\begin{eqnarray}
\Gamma_{\mbox{\scriptsize abs}}^{3, \mbox{\scriptsize 1-ph}} = {\pi \over \hbar^2}n(\omega_3)J_2(\omega_3), \\
\Gamma_{\mbox{\scriptsize em}}^{3, \mbox{\scriptsize 1-ph}} = {\pi \over \hbar^2}(n(\omega_3)+1)J_2(\omega_3),
\end{eqnarray}
where the superscript $``3"$ represents the spin transition $\ket{m_s = 0} \leftrightarrow \ket{m_s = -1}$. When 
$\omega_3 > 0$ the spin state $\ket{m_s = 0}$ is the lowest spin energy level and the absorption is defined by the transition 
$\ket{m_s = 0} \rightarrow \ket{m_s = -1}$. In the opposite case, i.e., when $\omega_3 < 0$, the damping rates can be written as the following
\begin{eqnarray}
\Gamma_{\mbox{\scriptsize abs}}^{3, \mbox{\scriptsize 1-ph}} = {\pi \over \hbar^2}n(|\omega_3|)J_2(|\omega_3|), \\
\Gamma_{\mbox{\scriptsize em}}^{3, \mbox{\scriptsize 1-ph}} = {\pi \over \hbar^2}(n(|\omega_3|)+1)J_2(|\omega_3|).
\end{eqnarray}
When $\omega_3 < 0$ the spin state $\ket{m_s=-1}$ is the lowest spin energy level. In this case, the absorption is defined by the transition $\ket{m_s = -1} \rightarrow \ket{m_s = 0}$. Figure.~\ref{fig:Figure2} shows the phonon-induced spin relaxation
rates associated with the ground triplet state of the NV$^{-}$ center as a function of the external magnetic field $B_0$. 
The absorption and emission damping rates associated with the transitions $\ket{m_s=0} \leftrightarrow \ket{m_s=-1}$ are shown only for the case $\omega_3 <0$. The total phonon-induced spin relaxation rates associated with one-phonon processes is defined as the sum of the absorption and emission transition rates of each process, and is given by
\begin{equation}
\Gamma{\mbox{\scriptsize 1-ph}} = \sum_{i=1}^{3}\left(\Gamma_{\mbox{\scriptsize abs}}^{i, \mbox{\scriptsize 1-ph}}  + \Gamma_{\mbox{\scriptsize em}}^{i, \mbox{\scriptsize 1-ph}} \right) = \sum_{i=1}^{3}A_i \coth\left({\hbar \omega_i \over 2 k_B T} \right). \label{Total-One-Phonon-Damping-Rate}
\end{equation}
This total phonon-induced spin relaxation rate will be relevant for the general solution associated with the populations of the spin states and the observable $\langle S_z(t)\rangle$ (see Section V and Eqs.\eqref{T1ArbitraryB0} and \eqref{LongitudinalRelaxationTime}). In addition, this transition rate, i.e., the sum of absorption and emission of all the transitions, is the rate that limits the coherence time $T_2$ \cite{Myers2017}. The parameters $A_i$ depend on the value of the spectral density function at the resonant frequencies, i.e., $A_1 = 2\pi J_1(\omega_1), A_2 = \pi J_2(\omega_2)$, and $A_3 = \pi J_2(|\omega_3|)$. \par

In the limit of continuous frequency, i.e., $\omega_k \rightarrow \omega$, we can introduce the following scaling for the linear spin-phonon coupling constants \cite{Weiss2008}:
\begin{eqnarray}
\lambda_{k,i} &\rightarrow & \lambda_{i}(\omega)  =  \lambda_{0i} \left({\omega \over \omega_D} \right)^{\nu}, \quad \quad 0 \leq \omega \leq \omega_D, \label{lambda(w)}
\end{eqnarray}
where $\lambda_i(\omega)$ is the one-phonon coupling constant for acoustic phonons, $\lambda_{0i} = \lambda_i(\omega_D)$ is the strength of the one-phonon coupling constant at the Debye frequency $\omega_D = \left(3/(4\pi n) \right)^{1/3}v_s$, where $n$ is the atom density, and $v_s$ is the speed of sound. For the diamond lattice the Debye frequency is given by $\omega_D/2\pi = 38.76$ THz \cite{Stedman1967}. The parameter $\nu$ is a phenomenological parameter that models the strength of the coupling for acoustic phonons and depends on the symmetry of the lattice. In the absence or presence of cubic symmetry we have $\nu=1/2$ or $\nu=3/2$, respectively \cite{Weiss2008}. For the NV$^{-}$ center in diamond we use the value $\nu = 1/2$, because of the presence of the color center with $C_{3v}$ symmetry that breaks the symmetry of the whole system (lattice and point defect). \par

We introduce the phonon density of states in a three-dimensional lattice $D(\omega) = \Omega \omega^2/(2\pi^2v_s^3)$, where $\Omega$ is the volume of a unit cell, $v_s = 1.2 \times 10^4$ m/s is the speed of sound in a diamond lattice, and $\omega_D$ is
the Debye frequency for the diamond lattice. In the limit of continuous frequency the spectral density functions can be written as $J_1(\omega) = [\lambda_x^2(\omega)+\lambda_y^2(\omega)]D(\omega)$ and $J_2(\omega) = [\lambda_{x'}^2(\omega)+\lambda_{y'}^2(\omega)]D(\omega)$. As a result, the parameters $A_i$ are given by
\begin{eqnarray}
A_1 &=& {\Omega \left(\lambda_{0x}^2 + \lambda_{0y}^2\right) \over \pi v_s^3 \omega_D}\left(2\gamma_s B \right)^3, \label{A1}\\
A_2 &=& {\Omega \left(\lambda_{0x'}^2 + \lambda_{0y'}^2\right) \over 2\pi v_s^3 \omega_D}\left(D + \gamma_s B \right)^3, \label{A2}\\
A_3 &=& {\Omega \left(\lambda_{0x'}^2 + \lambda_{0y'}^2\right) \over 2\pi v_s^3 \omega_D}\left(D - \gamma_s B \right)^3. \label{A3}
\end{eqnarray}
Therefore, the available number of phonons in the lattice, the density of phonon states, and the spin-phonon coupling constants will determine the intensity of each transition rate. In this context, the temperature is the control parameter in the laboratory that, at a quantum level, introduces available phonons that collectively act as a source of relaxation. At zero magnetic field, we have $A_1 = 0$ and $A_2 = A_3$. In the high-temperature regime, $k_B T \gg \hbar \omega_i$,
the one-phonon spin relaxation rates scales linearly with the temperature, i.e., $\Gamma^{i,\mbox{\scriptsize 1-ph}} \propto T$. In the opposite case, when $K_B T \ll \hbar\omega_i$, the one-phonon spin relaxation rates scales as a constant. \par

In the next section we introduce the second-order corrections to the phonon-induced spin relaxation rates given by the linear and bi-linear spin-phonon interaction terms.

\subsection{Two-phonon processes: acoustic phonons}

The second-order transition rate associated with the spin transition $\ket{m_s} \rightarrow \ket{m_s'}$ is defined as 
\begin{eqnarray}
\Gamma_{m_s \rightarrow m_s'} &=& \sum_{k,k'}\sum_{l,l'} \Gamma_{m_s,n_k,n_{k'}}^{m_s', n_{l},n_{l'}}, \quad m_s,m_s' = 0,\pm 1,
\label{Two-phonon-damping-rate}
\end{eqnarray}
where the sum is over all possible initial and final two-phonon modes, with $\ket{i} = \ket{m_s,n_k,n_{k'}}$ and $\ket{f} = \ket{m_s',n_l,n_{l'}}$ being the initial and final states, respectively. The transition rate inside the sum in Eq.~\eqref{Two-phonon-damping-rate} is given by the Fermi golden rule formula to second-order
\begin{eqnarray}
\Gamma_{m_s,n_k,n_{k'}}^{m_s', n_{l},n_{l'}} &=& {2 \pi \over \hbar^2}\left|V_{m_s,n_k,n_{k'}}^{m_s', n_{l},n_{l'}} \nonumber \right. \\ 
&&+\left.\sum_{m_s''=0,\pm 1}\sum_{p,p'}{V_{m_s',n_l,n_{l'}}^{m_s'',n_p,n_{p'}}  V_{m_s'',n_p,n_{p'}}^{m_s,n_k,n_{k'}}\over E_{m_s,n_k,n_{k'}} - E_{m_s'',n_p,n_{p'}} }\right|^2 \nonumber \\ 
&& \times \delta(\omega_{m_s',m_s}+ n_l \omega_l + n_{l'}\omega_{l'}- n_k \omega_k - n_{k'}\omega_{k'}), \nonumber\\ \label{FermiGoldenRule}
\end{eqnarray}
where $V_{i}^{j} = \bra{i} \hat{H}_{\mbox{\scriptsize s-ph}} \ket{j}$, $\ket{m_s''}$ is the spin state of the intermediate state, and $\ket{n_p}$,$\ket{n'_p}$ are the intermediate phonon states. The resonant frequencies of the system, i.e., $\omega_1 \sim 0-11.2 \; \mbox{GHz}$ and $\omega_{2,3} \sim 2.87-8.47 \; \mbox{GHz}$ are very low compared to the frequency of the acoustic phonons in diamond $\omega_{\mbox{\scriptsize acous}} \sim 0-10$ THz. Therefore, to second-order we assume that the most significant contribution comes from phonons that satisfy the frequency condition $\omega_{k,k'}\gg \omega_{m_s,m_s''}$.\par

\begin{figure}[t!]
\includegraphics[width= 1 \linewidth]{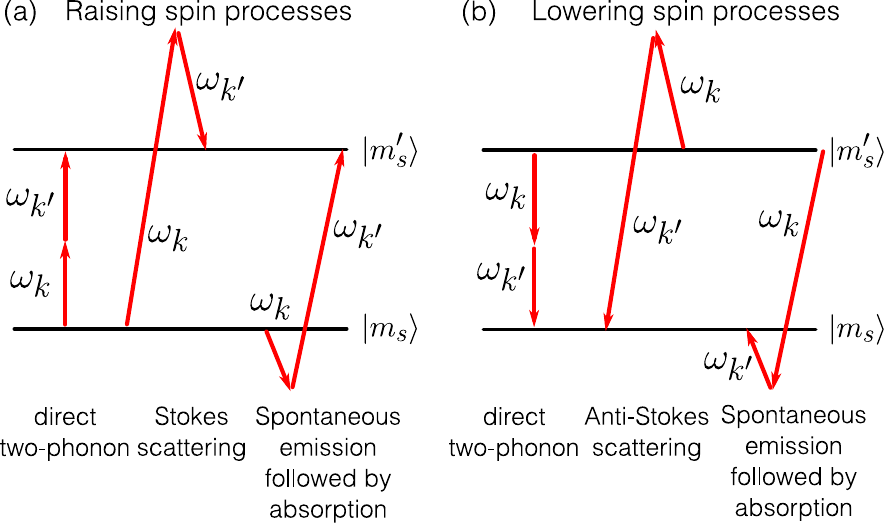}
\caption{The red arrows represent the absorption and emission of two phonons between two different spin states $\ket{m_s}$ and 
$\ket{m_s'}$. The direct two-phonon process is associated to the energy condition $\omega_k + \omega_{k'} = \omega_{m_s',m_s}$, where $\omega_{m_s',m_s} = \omega_{m_s'}-\omega_{m_s}$ is the frequency gap. The Stokes scattering
is associated to the energy condition $\omega_k - \omega_{k'} = \omega_{m_s',m_s}$.}
\label{fig:Figure3}
\end{figure}

We introduce four different types of two-phonon processes: two-phonon direct transition (Direct), Stokes transition (Stokes), anti-Stokes transition (anti-Stokes), and spontaneous emission followed by absorption (Spont), see Fig.~\ref{fig:Figure3}. 
The direct two-phonon transition is characterized by the frequency condition $\omega_k + \omega_{k'} = \omega_{m_s',m_s}$ and its absorption and emission relaxation rates are given by
\begin{eqnarray}
\Gamma_{m_s \rightarrow m_s'}^{\mbox{\scriptsize abs, Direct}} &=& \sum_{k,k'} \Gamma_{m_s,n_k,n_{k'}}^{m_s', n_k-1,n_{k'}-1}, \\
\Gamma_{m_s' \rightarrow m_s}^{\mbox{\scriptsize em, Direct}} &=& \sum_{k,k'} \Gamma^{m_s,n_k+1,n_{k'}+1}_{m_s', n_k,n_{k'}}. 
\end{eqnarray}
On the other hand, we have the Stokes and Anti-Stokes transitions which are characterized by the frequency condition $\omega_k - \omega_{k'} = \omega_{m_s',m_s}$ and are given by
\begin{eqnarray}
\Gamma_{m_s \rightarrow m_s'}^{\mbox{\scriptsize Stokes}} &=& \sum_{k,k'} \Gamma_{m_s,n_k,n_{k'}}^{m_s', n_k-1,n_{k'}+1}, \\
\Gamma_{m_s' \rightarrow m_s}^{\mbox{\scriptsize Anti-Stokes}} &=& \sum_{k,k'} \Gamma^{m_s,n_k-1,n_{k'}+1}_{m_s', n_k,n_{k'}}.
\end{eqnarray}
For the spontaneous emission followed by absorption process we define
\begin{eqnarray}
\Gamma_{m_s \rightarrow m_s'}^{\mbox{\scriptsize abs,Spont}} &=& \sum_{k,k'} \Gamma_{m_s,n_k,n_{k'}}^{m_s', n_k+1,n_{k'}-1},  \\
\Gamma_{m_s' \rightarrow m_s}^{\mbox{\scriptsize em, Spont}} &=& \sum_{k,k'} \Gamma^{m_s,n_k+1,n_{k'}-1}_{m_s', n_k,n_{k'}}. 
\end{eqnarray}
For acoustic phonon modes, i.e., phonons with a linear dispersion relation $\omega_k = v|\vec{k}|$, we can use the Debye model in order to represent two-phonon processes. In order to study the spin-relaxation rate as a function of the dimension of the system, we introduce the density of phonon states for a $d$-dimensional lattice.
\begin{equation}
D(\omega) = D_0 \left({\omega \over \omega_D}\right)^{d-1}, \quad \quad 0 \leq \omega \leq \omega_D. \label{DOS-d}
\end{equation}
Here, $\omega_D$ is the Debye frequency for the diamond lattice, $D_0>0$ is a normalization constant, and $d=1,2,3$ is the dimension of the lattice. We can introduce the following scaling for the quadratic spin-phonon coupling constant for the acoustic phonon modes \cite{Weiss2008}
\begin{eqnarray}
\lambda_{k k',i} &\rightarrow& \lambda_{i}(\omega,\omega') =  \lambda_{00i} \left({\omega \over \omega_D} \right)^{\nu} \left({\omega' \over \omega_D} \right)^{\nu},
\end{eqnarray}
where $\lambda_{i}(\omega,\omega')$ is the two-phonon coupling constant for acoustic phonons, $\lambda_{00i} = \lambda_{i}(\omega_D,\omega_D)$ is the strength of the two-phonon coupling constant at the Debye frequency $\omega_D$, and $\nu > 0$ is a phenomenological factor that models the spin-phonon coupling in the acoustic regime. \par

Using the second-order Fermi golden rule given in Eq.~\eqref{Final-Spin-Phonon-Hamiltonian} and only considering acoustic phonons, we obtain the following absorption and emission transition rates
\begin{eqnarray}
\Gamma_{m_s \rightarrow m_s'}^{\mbox{\scriptsize abs}} &=& \Gamma_{m_s \rightarrow m_s'}^{\mbox{\scriptsize abs, Direct}} +
\Gamma_{m_s \rightarrow m_s'}^{\mbox{\scriptsize Stokes}} + \Gamma_{m_s \rightarrow m_s'}^{\mbox{\scriptsize abs,Spont}},\\
\Gamma_{m_s \rightarrow m_s'}^{\mbox{\scriptsize em}} &=& \Gamma_{m_s \rightarrow m_s'}^{\mbox{\scriptsize em, Direct}} +
\Gamma_{m_s \rightarrow m_s'}^{\mbox{\scriptsize Anti-Stokes}} + \Gamma_{m_s \rightarrow m_s'}^{\mbox{\scriptsize em,Spont}},
\end{eqnarray}
where each transition rate is defined as 
\begin{eqnarray}
\Gamma^{\mbox{\scriptsize process}}_{m_s \rightarrow m_s'} &=& a^{\mbox{\scriptsize process}}_{m_s,m_s'}(x_D)T^{4\nu+2d-3}+b^{\mbox{\scriptsize process}}_{m_s,m_s'}(x_D)T^{4\nu+2d-2}\nonumber \\
&+&c^{\mbox{\scriptsize process}}_{m_s,m_s'}(x_D)T^{4\nu+2d-1},
\end{eqnarray}
where process = \{Direct, Stokes, Anti-Stokes, Spont\}, $x_D = \hbar \omega_D / k_B T$ is a dimensionless parameter, $T$ is the temperature, and the coefficients $a^{\mbox{\scriptsize process}}_{m_s,m_s'}$, $b^{\mbox{\scriptsize process}}_{m_s,m_s'}$, and 
$c^{\mbox{\scriptsize process}}_{m_s,m_s'}$ are given in Appendix A. Using $\nu = 1/2$ and $d = 3$, we obtain the following total two-phonon spin relaxation rate
\begin{eqnarray}
\Gamma_{\mbox{\scriptsize 2-ph}} &=& 
\sum_{m_s \neq m_s'}\left(\Gamma_{m_s \rightarrow m_s'}^{\mbox{\scriptsize abs}} +\Gamma_{m_s \rightarrow m_s'}^{\mbox{\scriptsize abs}}\right) \nonumber \\
&=& A_5 T^5 + A_6 T^6 + A_7 T^7.
\end{eqnarray}
This total spin relaxation rate will be relevant for the general solution associated with the physical observable $\langle S_z(t)\rangle$ (see Section V and Eq.~\eqref{LongitudinalRelaxationTime}). In Table I, we have shown the different temperature dependence of the spin relaxation rate associated with two-phonon processes in the acoustic limit. We observe that the symmetry of the lattice $\nu$ and the dimension of the system $d$ determine the temperature response of the spin-lattice relaxation dynamics of the system at high temperatures. \par

In summary, by only considering the contribution of acoustic phonons to first and second-order, we see three
different temperature scalings of the form ($T^s, T^{s+1}, T^{s+2}$), where $s = 4\nu + 2d-3$. We observe 
$1/T_1 \propto T^s$ for a linear second-order Raman-like scattering, $1/T_1 \propto T^{s+2}$ for a quadratic first-order Raman-like scattering, and $1/T_1 \propto T^{s+1}$ for the mixed term between the linear and quadratic
contributions to second order.

\begin{table}[]
\centering
\caption{The table shows the expected temperature dependence
of linear and bi-linear spin-phonon interactions considered
to first and second order. The bi-linear term to second order is
not considered as it is a small contribution. When both linear
and bi-linear terms are considered a mixed term appears only to
second-order. Last column indicates the power factor for relaxation
of the form $T^s$ in three dimensions and for a non-cubic
lattice ($\nu = 1/2$).}
\label{my-label}
\begin{tabular}{|c|c|c|c|}
\hline
\multicolumn{1}{|c|}{Hamiltonian} & \multicolumn{1}{c|}{First-order} & \multicolumn{1}{c|}{Second-order} & \multicolumn{1}{c|}{
$\begin{array}{c}
 d=3 \\
 \nu = 1/2
\end{array}$} \\ \hline
      $\displaystyle{\hat{H} = \sum_{k,i} \lambda_{ki} \hat{x}_k}$         &      $\displaystyle{\coth\left({\hbar \omega \over k_B T} \right)}$                 &     $T^{4\nu + 2d -3}$                   &    $T^5$              \\ \hline
      $\displaystyle{\hat{H} = \sum_{k,k',i} \lambda_{kk',i} \hat{x}_k \hat{x}_{k'}}$                &       $T^{4\nu + 2d -1}$                &                     &    $T^7$                   \\ \hline
       \mbox{Mixed term}                &                     &          $T^{4\nu + 2d -2}$                               &          $T^6$             \\ \hline
\end{tabular}
\end{table}

\subsection{Two-phonon processes: quasi-localized phonons}
Quasi-localized phonons, or vibrational resonances between a single-color-center and lattice vibrations, are good
candidates for dissipative processes due to the strong electron-phonon coupling. The NV$^{-}$ center has a strong
electron-phonon coupling associated with vibrational resonances, with a continuum of vibrational modes centered at
$\omega_{\mbox{\scriptsize  res}} = 65$ meV, and a full width at half-maximum of about $\Delta = 32$ meV as regularly observed in the phonon-sideband of the NV fluorescence spectrum under optical excitation \cite{Alkauskas2014}. Because of the small zero-field splitting constant induced by spin-spin interaction ($D/2\pi = 2.87$ GHz or 
$\hbar D = 0.012$ meV), we have $\omega_{\mbox{\scriptsize  res}} \gg D$, and therefore, these high-energy phonons can only be present in a two-phonon process associated with the condition $\omega_k-\omega_{k'} = \omega_i$ ($\omega_k \approx \omega_{k'}$).
Strong interactions with high energy phonons can be introduced in Orbach-type processes \cite{Redman1991}. It is shown experimentally that different NV$^{-}$ center samples have an activation energy of 73 meV \cite{Jarmola2012}, which is close to the vibrational resonance frequency $\omega_{\mbox{\scriptsize  res}} = 65$ meV. In our formalism, quasi-localized phonons can be phenomenologically modeled by a Lorentzian spectral density function of the form \cite{Thorwart2000, Ariel2016}
 \begin{equation}
J_{\mbox{\scriptsize  Loc}}(\omega)  = {J_{\mbox{\scriptsize  Loc}} \over \pi}{{1 \over 2}\Delta \over \left(\omega - \omega_{\mbox{\scriptsize  loc}} \right)^2 +  \left({1 \over 2}\Delta \right)^2},  
\hspace{0.5 cm} 0<\omega<\omega_{\mbox{\scriptsize max}}. \label{JLorentz}
\end{equation}
In this equation, $J_{\mbox{\scriptsize  Loc}}$ is the coupling strength, $\Delta$ is a characteristic bandwidth, $\hbar\omega_{\mbox{\scriptsize max}} = 168$ meV is the maximum phonon energy in a diamond lattice \cite{BookZaitsev2011}, and $\omega_{\mbox{\scriptsize  loc}}$ is the frequency of the localized phonon mode. As a simpler
model we can consider the interaction with only one quasi-localized phonon mode ($\Delta \rightarrow 0$)
\begin{eqnarray}
\lambda_{k,i} &=&  \lambda_{i,\mbox{\scriptsize loc}} \delta(\omega-\omega_{\mbox{\scriptsize  loc}}), \label{LocalizedInteraction}
\end{eqnarray}
where $\lambda_{i,\mbox{\scriptsize loc}}$ is the coupling strength. Using the above equation and calculating the second-order 
transition rate induced by the linear spin-phonon interaction, we can obtain the following relaxation rate associated with quasi-localized phonons
\begin{eqnarray} 
\Gamma_{\mbox{\scriptsize loc}} &=&  A_4 \left(1 + n( \omega_{\mbox{\scriptsize  loc}}) \right) n(\omega_{\mbox{\scriptsize  loc}}) \nonumber \\
&\approx&  {A_4 \over e^{\hbar\omega_{\mbox{\scriptsize  loc}} /k_B T}-1},                 
\end{eqnarray} 
where $A_4$ is a constant of units of frequency. The approximation $\left(1 + n( \omega_{\mbox{\scriptsize  loc}}) \right) n(\omega_{\mbox{\scriptsize  loc}}) \approx n(\omega_{\mbox{\scriptsize  loc}})$ is valid for temperatures
below T = 300 K. For such temperatures, the mean number of phonons is low, $n(\omega_{\mbox{\scriptsize  loc}}) \approx 0.1$, therefore we can write $(1+n)n \approx n + \mathcal{O}(n^2)$. \par

In the next section we derive the spin-lattice relaxation dynamics using the quantum
master equation.

\section{Spin-lattice relaxation dynamics}
In this section, we present the general equation associated with the spin-lattice relaxation dynamics of the ground
triplet state of the NV$^{-}$ center. We use the non-Markovian quantum master equation \cite{Breuer2002} for the reduced density operator $\hat{\rho}(t) = \mbox{Tr}_{\mbox{\scriptsize ph}}\left(\hat{\rho}_{\mbox{\scriptsize NV+ph}}\right)$. We assume that the initial state at time $t_0$ is given by the uncorrelated state $\hat{\rho}_{\mbox{\scriptsize NV+ph}}(t_0) = \hat{\rho}_{\mbox{\scriptsize NV}}(t_0) \otimes \hat{\rho_{\mbox{\scriptsize ph}}}(t_0)$ (Born approximation), and that the phonon bath is in thermal equilibrium. In the weak-coupling limit, and using the spin-phonon Hamiltonian given in Eq.~\eqref{Final-Spin-Phonon-Hamiltonian}, we obtain
\begin{eqnarray}
\dot{\hat{\rho}} &=& {1 \over i\hbar} [\hat{H}_{\mbox{\scriptsize NV}},\hat{\rho}] + \mathcal{L}_{\mbox{\scriptsize 1-ph}}\hat{\rho}  +  \mathcal{L}_{\mbox{\scriptsize 2-ph}} \hat{\rho} + \mathcal{L}_{\mbox{\scriptsize mag}} \hat{\rho},  \label{QME}
\end{eqnarray} 
where the first term in Eq.~\eqref{QME} describes the free dynamics induced by the NV$^{-}$ center Hamiltonian [Eq.~\eqref{SpinHamiltonian}]. The second and third terms are given by 
\begin{eqnarray}
\mathcal{L}_{\mbox{\scriptsize 1-ph}}  \hat{\rho} &=& \sum_{i=1}^{3}\left[\Gamma_{\mbox{\scriptsize abs}}^{i, \mbox{\scriptsize 1-ph}}  \mathcal{D}[L_+^{i}]\hat{\rho} + \Gamma_{\mbox{\scriptsize em}}^{i, \mbox{\scriptsize 1-ph}} \mathcal{D}[L_-^{i}]\hat{\rho} \right], \label{L1} \\
\mathcal{L}_{\mbox{\scriptsize 2-ph}}  \hat{\rho} &=& \sum_{i=1}^{3}\left[\Gamma_{\mbox{\scriptsize abs}}^{i, \mbox{\scriptsize 2-ph}} \mathcal{D}[L_+^{i}]\hat{\rho} + \Gamma_{\mbox{\scriptsize em}}^{i, \mbox{\scriptsize 2-ph}} \mathcal{D}[L_-^{i}]\hat{\rho} \right], \label{L2}
\end{eqnarray}
which describe the dissipative spin-lattice dynamics induced by one-phonon and two-phonon processes, with the
index $i=1,2,3$ representing the spin transitions of the system (see Fig.~\ref{fig:Figure1}). In Eqs.~\eqref{L1} and \eqref{L2} we have defined the Lindblad super-operator $\mathcal{D}[\hat{O}] \hat{\rho} = \hat{O} \hat{\rho} \hat{O}^{\dagger} - {1 \over 2}\{\hat{O}^{\dagger} \hat{O}, \hat{\rho}\}$ and the spin operators 
\begin{eqnarray}
L_+^{1} &=& \ket{m_s=1}\bra{m_s=-1} = \left( L_{-}^{1} \right)^{\dagger}, \\
L_+^{2} &=& \ket{m_s=1}\bra{m_s=0}= \left( L_{-}^{2} \right)^{\dagger}, \\
L_+^{3} &=& \ket{m_s=-1}\bra{m_s=0}= \left( L_{-}^{3} \right)^{\dagger}.
\end{eqnarray}
The last term in Eq.~\eqref{QME} is an extra term that describes a phenomenological dynamics induced by magnetic impurities,
and is given by
\begin{flalign}
&\mathcal{L}_{\mbox{\scriptsize mag}} \hat{\rho}= -{1 \over 4}\Gamma_{\mbox{\scriptsize mag}}\sum_{i=x,y,z} \left[S_i,[S_i, \hat{\rho}(t)]\right], 
\end{flalign}
where $\Gamma_{\mbox{\scriptsize mag}}$ is the magnetic relaxation rate induced by an isotropic magnetic noise \cite{Avron2015}, and $S_i$ are the Pauli matrices for $S=1$. From previous works, it is expected that the parameter $\Gamma_{\mbox{\scriptsize mag}}$ will proportionally depend on the concentration of neighboring NV$^{-}$ centers \cite{Jarmola2012} and temperature. Therefore, $\Gamma_{\mbox{\scriptsize mag}}$ is a sample-dependent parameter that models magnetic impurities. The exact temperature dependence of $\Gamma_{\mbox{\scriptsize mag}}$ is beyond the scope of this work, but we expect it to change as temperature reaches $T_{\mbox{\scriptsize gap}} = \hbar D/k_B \approx 0.14$ K. In addition, in this work we neglect the effect of electric field fluctuations. This is relevant for experiments that involve optical illumination and read-out of the electronic states \cite{Myers2017}. \par

Now, we study the longitudinal relaxation rate at low and high temperatures. In the low-temperature limit we also investigate the effect of magnetic field on the longitudinal relaxation rate.

\section{Discussion}

\subsection{Low-temperature limit}
\begin{figure}[t!]
\includegraphics[width= 1 \linewidth]{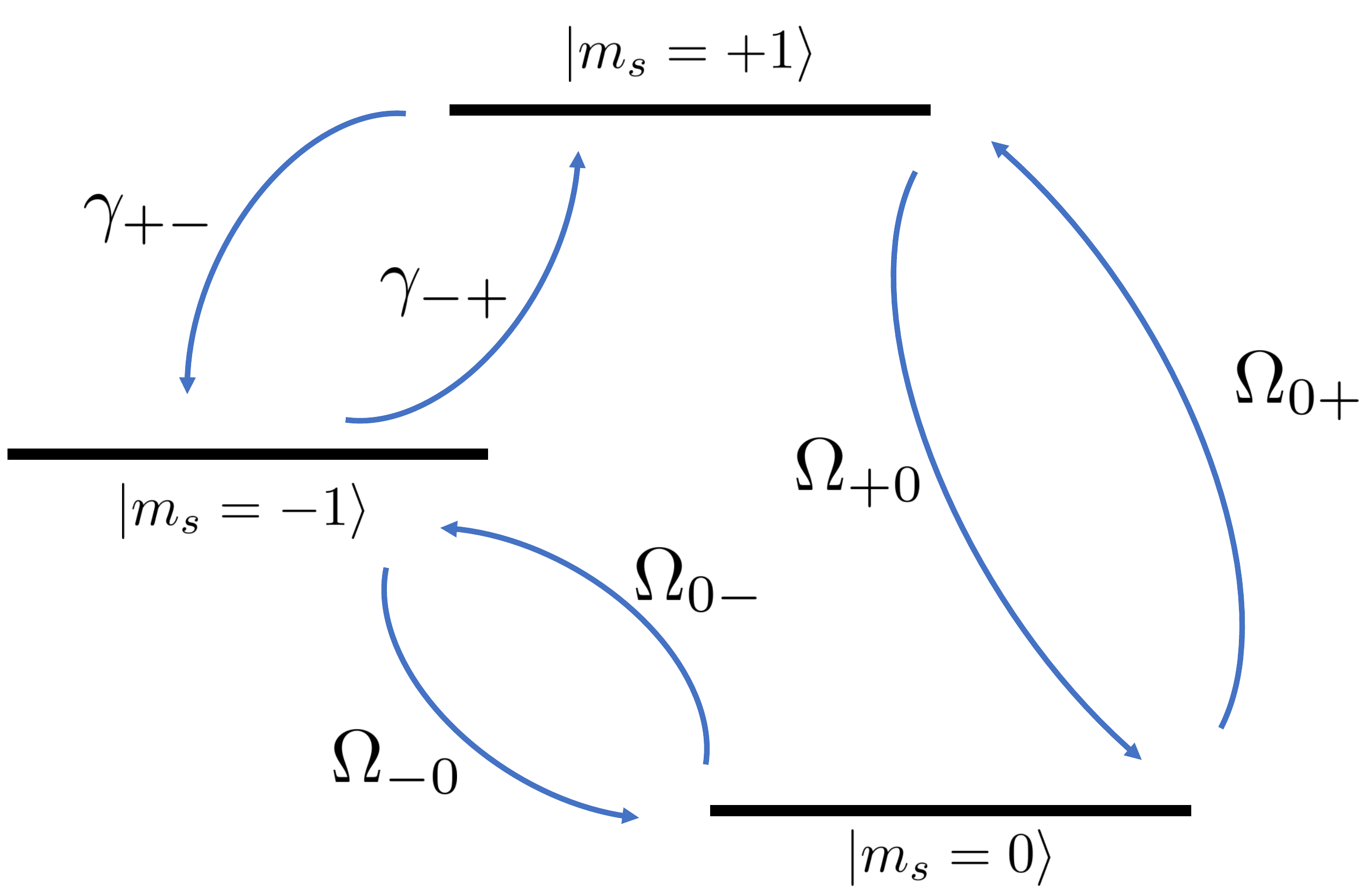}
\caption{Direct relaxation rates induced by one-phonon processes. The spin populations associated with the spin states $\ket{m_s = 0, \pm 1}$ are modified by the absorption ($\gamma_{-+}, \Omega_{0-}, \Omega_{0+}$) and emission rates ($\gamma_{+-}, \Omega_{-0}, \Omega_{+0}$). For magnetic fields $\gamma_s B_0 > D$ ($B_0 > 1000$ G), the state $\ket{m_s=-1}$ is the lowest energy state and the role of $\Omega_{0-}$ and $\Omega_{-0}$ are exchanged.}
\label{fig:Figure6}
\end{figure}

In this section we discuss the low-temperature limit (below 1 K) associated to the spin-lattice relaxation dynamics
of the ground state of the NV$^{-}$ center in diamond. For low temperatures, only one-phonon processes contribute to
the transition rates. Therefore, we can deduce the spin-lattice dynamics from the quantum master equation by setting $\mathcal{L}_{\mbox{\scriptsize 2-ph}} \hat{\rho} = 0$. From Eq.~\eqref{QME} we can find the dynamics
of the spin populations $p_1 = \bra{m_s=1}\hat{\rho}\ket{m_s=1}$, $p_2 = \bra{m_s=0}\hat{\rho} \ket{m_s=0}$, and $p_3 = \bra{m_s=-1}\hat{\rho}\ket{m_s=-1}$. For an arbitrary magnetic field $B_0$ along the $z$ axis, using $\Gamma_{\mbox{\scriptsize mag}} = 0$, and considering only one-phonon processes, the equations at low temperatures are given by
\begin{eqnarray}
{\mbox{d} p_1 \over \mbox{d}t} &=& - \left(\gamma_{+-}+\Omega_{+0}\right) p_1 + \Omega_{0+} p_2 + \gamma_{-+} p_3, \label{p1}\\
{\mbox{d} p_2 \over \mbox{d}t} &=& - \left(\Omega_{0+}+\Omega_{0-} \right) p_2 +  \Omega_{+0}p_1 +  \Omega_{-0}p_3, \label{p2}\\
{\mbox{d} p_3 \over \mbox{d}t} &=& - \left(\Omega_{0-}+\gamma_{-+} \right) p_3 + \gamma_{+-} p_1 + \Omega_{0-} p_2, \label{p3} 
\end{eqnarray}
where the direct relaxation rates between the spin states are given by $\gamma_{+-} = A_1(1+n_1)$, $\gamma_{-+} = A_1 n_1$, $\Omega_{+0} = A_2(1+n_2)$, $\Omega_{0+} = A_2n_2$, $\Omega_{-0} = A_3(1+n_3)$, and $\Omega_{0-} = A_3 n_3$ (see Fig.~\ref{fig:Figure6}),  where $n_i = \left[\mbox{exp}(\hbar \omega_i / k_B T)-1\right]^{-1}$ the mean number of phonons at thermal equilibrium. Here, $\omega_1 = 2\gamma_s B_0$, $\omega_2 = D + \gamma_s B_0$, and $\omega_3 = D - \gamma_s B_0$ are the resonant frequencies associated with the spin energy levels. The $A_i$ parameters are defined in Eqs.~\eqref{A1}-\eqref{A3} and are estimated as a function of the magnetic field $B_0$ in the next section [see Eqs.~\eqref{A1-estimated}-\eqref{A3-estimated}]. For experiments in quantum information processing and magnetometry these direct relaxation rates plays a fundamental role. \par

In the following we obtain the longitudinal relaxation rate for the physical observables $\langle S_z^2(t)\rangle$ and $\langle S_z(t)\rangle$ at different magnetic field regimes. However, this model can be used to determine any other physical observable, for instance, direct relaxation rates between spin states and its magnetic field and temperature dependence. \par

\subsubsection{Zero magnetic field}
At zero magnetic field ($B_0 = 0$) and neglecting the effect of strain, the spin states $\ket{m_s = 1}$ and $\ket{m_s = -1}$ are degenerate (see Fig.~\ref{fig:Figure1}). As a consequence, the emission and absorption rates associated with the spin transitions $\ket{m_s=0} \leftrightarrow \ket{m_s = 1}$ and $\ket{m_s=0} \leftrightarrow \ket{m_s = -1}$ are equal. 

\begin{figure}[t!]
\includegraphics[width= 1 \linewidth]{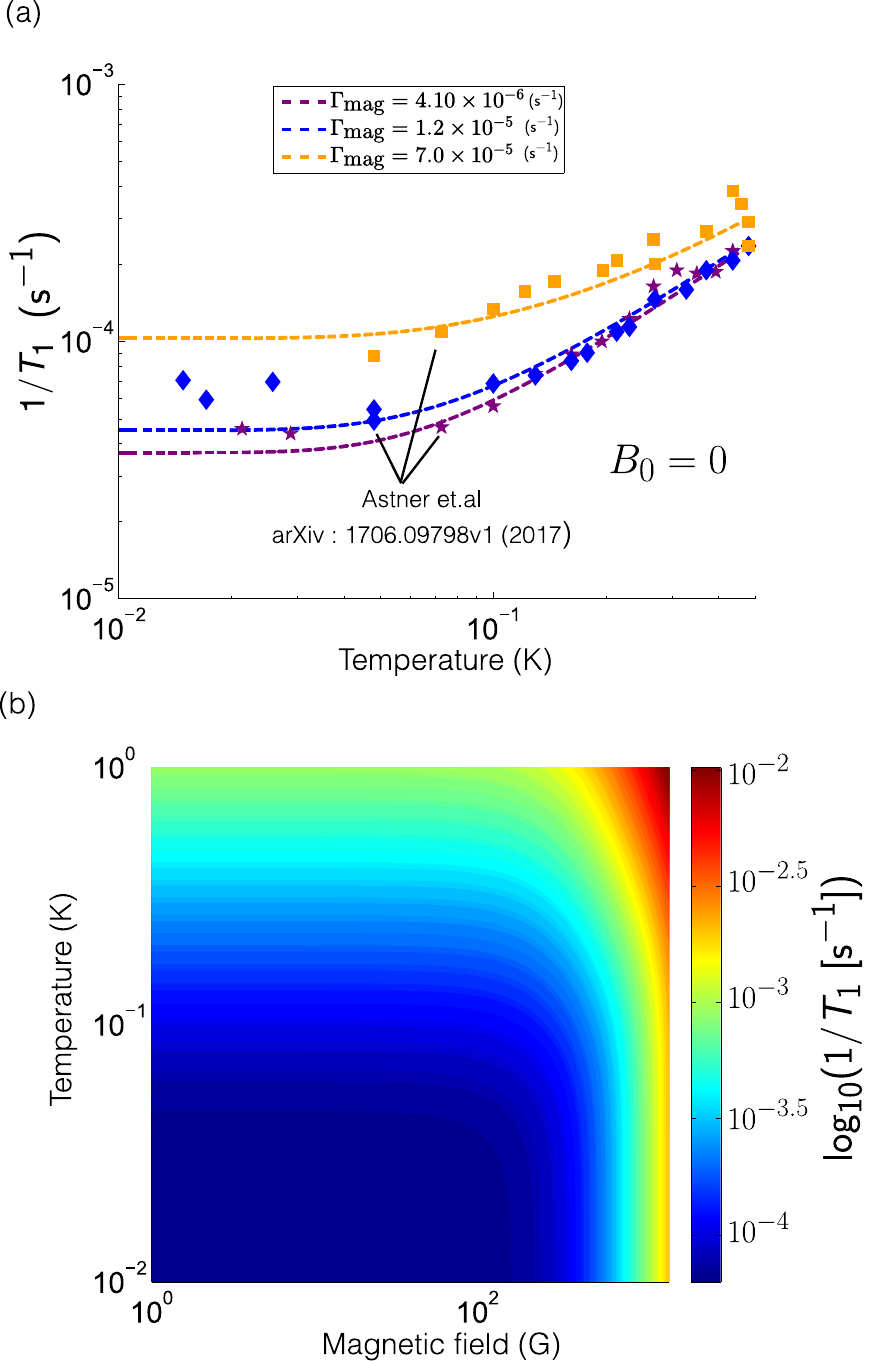}
\caption{(a) Relaxation rate of $\langle S_z^2(t)\rangle$ at zero magnetic field. The symbols represent experimental spin relaxation rates measured at low temperatures (below 1 K) for different NV-samples \cite{Astner2017}. The dotted lines represent the theoretical fit given by $1/T_1 = \Gamma_0(1+3\bar{n}) + \Gamma_{\mbox{\scriptsize mag}}$. We observe that at low temperatures, the relevant contribution comes from the emission of a phonon and the magnetic noise induced by the environment. (b) Two-dimensional parameter plot of the longitudinal relaxation rate of $\langle S_z(t) \rangle$ in logarithm scale at magnetic fields ranging from 0 to 1500 G, temperature ranging from 10 mK to 1 K, and $\Gamma_{\mbox{\scriptsize}} = 0$.}
\label{fig:Figure4}
\end{figure}

Therefore, the system can be modeled as a simple two-level system with the degenerate excited states described by $\ket{m_s = \pm 1}$. In addition, the transition rate between $\ket{m_s = \pm 1}$ vanishes if we neglect the effect of electric field fluctuations \cite{Myers2017}. In such scenario, the absorption and emission rates are given by $\Gamma_{\mbox{\scriptsize abs}} = \Gamma_0 \bar{n}$ and $\Gamma_{\mbox{\scriptsize em}} = \Gamma_0 (\bar{n}+1)$, respectively, where $\bar{n} = \left[\mbox{exp}(\hbar D / k_B T)-1\right]^{-1}$ is the mean number of phonons at the zero-field splitting frequency $D/2\pi = 2.87 \; \mbox{GHz}$. The parameter $\Gamma_0$ is obtained from Eqs.~\eqref{A2} and \eqref{A3}) for $B_0=0$ and is given by
\begin{equation}
\Gamma_0 = {\Omega D^3(\lambda_{0x'}^2+\lambda_{0y'}^2) \over 2\pi v_s^3 \omega_D}. \label{Gamma0}
\end{equation}
From Eqs.~\eqref{p1}-\eqref{p3}, we obtain
\begin{eqnarray}
{\mbox{d} p_1 \over \mbox{d}t} &=& \Gamma_0(1 + \bar{n})p_1 + \Gamma_0 \bar{n} p_2,\\
{\mbox{d} p_2 \over \mbox{d}t} &=& -2\Gamma_0 \bar{n} p_2 + \Gamma_0(1+\bar{n}) p_1 + \Gamma_0(1+\bar{n})p_3,\\
{\mbox{d} p_3 \over \mbox{d}t} &=& \Gamma_0(1 + \bar{n})p_1 + \Gamma_0 \bar{n} p_3.
\end{eqnarray}
Using $\langle S_z^2(t)\rangle = p_1(t) + p_3(t)$ and $p_1(t)+p_2(t)+p_3(t) = 1$ we obtain
\begin{eqnarray}
{\mbox{d} \langle S_z^2(t)\rangle \over \mbox{d}t} &=& -\Gamma_0(1 + 3\bar{n})\langle S_z^2(t)\rangle + 2\Gamma_0 \bar{n},\\
{\mbox{d} p_2 \over \mbox{d}t} &=&  -\Gamma_0(1 + 3\bar{n}) p_2(t) + \Gamma_0 (1+\bar{n}).
\end{eqnarray}
Using arbitrary initial conditions $p_i(0) = p_{i0}$ ($i=1,2,3$), we
have
\begin{eqnarray}
\langle S_z^2(t)\rangle &=& \langle S_z^2(T)\rangle_{\mbox{\scriptsize st}} - \left(\langle S_z^2(T)\rangle_{\mbox{\scriptsize st}} -p_{10} -p_{30} \right)e^{-\Gamma_0(1 + 3\bar{n})t}, \nonumber \\
p_2(t) &=& \left(p_2(T)\right)_{\mbox{\scriptsize st}} - \left(\left(p_2(T)\right)_{\mbox{\scriptsize st}} -p_{20} \right)e^{-\Gamma_0(1 + 3\bar{n})t},
\end{eqnarray}
where the steady states are given by
\begin{eqnarray}
\langle S_z^2(T)\rangle_{\mbox{\scriptsize st}} &=& {2 \over e^{\hbar D/ k_B T}+2}, \\
\left(p_2(T)\right)_{\mbox{\scriptsize st}} &=& {e^{\hbar D/ k_B T} \over e^{\hbar D/ k_B T}+2}.
\end{eqnarray}
Therefore, the phonon-induced spin relaxation rate associated with $\langle S_z^2(t)\rangle$ and $p_2(t)$ (ground state population) are given by $\Gamma_0(1 + 3\bar{n})$, where $\Gamma_0 = 3.14 \times 10^{-5}$ s$^{-1}$ \cite{Astner2017}. This is
consistent with the longitudinal relaxation rate recently measured and estimated by \textit{ab initio} methods
in Ref.~\cite{Astner2017} (see Fig.~\ref{fig:Figure4}a). Using Eq.~\eqref{Gamma0} and assuming $\lambda_{0x'} \approx \lambda_{0y'} \approx \lambda_{0x} \approx \lambda_{0y}$, we estimate $\lambda_{0x'}$ to be approximately $3.97$ meV. With this approximation for the $\lambda_0$ factors and combining Eqs.~\eqref{A1}-\eqref{A3} with Eq.~\eqref{Gamma0}, we can estimate the following magnetic field dependence for the one-phonon spin relaxation rates
\begin{eqnarray}
A_1 &\approx& 2 \Gamma_0 \left({2\gamma_s B_0 \over D}\right)^3, \label{A1-estimated} \\
A_2 &\approx& \Gamma_0 \left[{(D+\gamma_s B_0) \over D}\right]^3,\label{A2-estimated} \\
A_3 &\approx& \Gamma_0 \left[{(D-\gamma_s B_0) \over D}\right]^3. \label{A3-estimated}
\end{eqnarray}
Note that $\langle S_z(t)\rangle$ is zero as the states $\ket{m_s = +1}$ and $\ket{m_s = -1}$ are degenerate at zero magnetic field. In the next section we introduce the effect of low magnetic field on the longitudinal relaxation rate associated with $\langle S_z(t)\rangle$.

\subsubsection{Low magnetic field}
We define the limit of low magnetic fields when $\gamma_s B_0 \ll D$ so that $n(D+\gamma_s B_0) \approx n(D-\gamma_s B_0) \approx \bar{n}$. By considering one-phonon processes, we obtain the following set of equations
\begin{eqnarray}
{\mbox{d} \langle S_z^2(t)\rangle \over \mbox{d}t} &=& -\Gamma_0(1+3\bar{n})\langle S_z^2(t)\rangle + 
3 \epsilon \Gamma_0(1+\bar{n})\langle S_z(t)\rangle \nonumber \\ 
&& +2\Gamma_0 \bar{n},\\
{\mbox{d} \langle S_z(t)\rangle \over \mbox{d}t} &=& -\left[\Gamma_B n_B + 3\epsilon \Gamma_0(1+3\bar{n}) \right] \langle S_z^2(t)\rangle + 6\epsilon \Gamma_0 \bar{n} \nonumber \\
&& - \left[\Gamma_B(1+ 2 n_B) + \Gamma_0(1+\bar{n}) \right]\langle S_z(t)\rangle,
\end{eqnarray}
where $\epsilon = \gamma_s B_0 / D \ll 1$ is a pertubative dimensionless parameter, $\Gamma_B \approx \Gamma_0 (2\gamma_s B_0/D)^3$, and $n_B = \left[\mbox{exp}(2\hbar \gamma_s B_0   / k_B T)-1\right]^{-1}$ is the mean number of phonons
at the resonant frequency $\omega_1 = 2\gamma_s B_0$. In addition, the mean number of phonons satisfies 
$n_B \gg \bar{n}$ due to the condition $\gamma_s B_0 \ll D$. 

At low magnetic fields, the longitudinal relaxation rate associated with $\langle S_z(t)\rangle$ is given by
\begin{equation}
{1 \over T_1} \approx  2\Gamma_0(1 + 2\bar{n}) + \Gamma_B(1 + 2n_B). \label{T1LowB}
\end{equation}
The steady states satisfy the relation
\begin{equation}
{ \langle S_z^2(T) \rangle_{\mbox{\scriptsize st}} \over  \langle S_z(T) \rangle_{\mbox{\scriptsize st}}} = 
{\Gamma_0 (1+\bar{n})+\Gamma_B(1+2 n_B) \over n_B \Gamma_B}.
\end{equation}
In the next section we obtain the longitudinal relaxation rate associated with $\langle S_z(t)\rangle$ for arbitrary values of the magnetic field $B_0$.

\subsubsection{Arbitrary magnetic field values}
At non-zero magnetic fields, the spin states $\ket{m_s=-1}$ and $\ket{m_s=1}$ are split due to the Zeeman interaction (see Fig.~\ref{fig:Figure1}). This implies that the system can be modeled as a dissipative three-level system consisting of the spin states 
$\ket{m_s=0}$ and $\ket{m_s = \pm 1}$. From Eqs.~\eqref{p1}-\eqref{p3}, the dynamics for the longitudinal spin component is given by
\begin{equation}
{\mbox{d}^2 \langle S_z(t)\rangle \over \mbox{d}t^2} + {1 \over T_1}{\mbox{d} \langle S_z(t)\rangle \over \mbox{d}t} 
+\omega^2\langle S_z(t)\rangle = A_0,
\end{equation}
where the parameters are given by
\begin{eqnarray}
{1 \over T_1} &=& A_1 (1+ 2 n_1) + A_2(1+2n_2)+ A_3(1+2n_3), \label{T1ArbitraryB0} \\
\omega^2 &=& {1 \over 2}\left\{A_1[A_3(1+n_3)-A_2(1+n_2)]-A_2^2n_2(3+n_2) \right. \nonumber \\ 
         & & \left. -A_3^2 n_3(3+n_3)+A_2 A_3(2+n_2+n_3+4n_2 n_3)\right\}, \nonumber \\
A_0 &=& {1 \over 2}\left[2A_1(A_2+A_3)+A_2^2(1+n_2)^2 \right. \nonumber \\
    & & \left. +2A_2A_3(n_2-n_3)-A_3^2(1+n_3)^2 \right].       
\end{eqnarray}
We observe that the damping rate $1/T_1$ is given by the total one-phonon spin relaxation rate given in Eq.~\eqref{Total-One-Phonon-Damping-Rate}. The general solution is that of a driven damped harmonic oscillator, where the longitudinal relaxation rate is given
by
\begin{equation}
{1 \over T_1} \approx {\Gamma_0 \over D^3}\omega_1^3 n_1 + {\Gamma_0 \over D^3}\left[\omega_2^3(1+2n_2)+ \omega_3^3(1+2n_3)\right],
\end{equation}
where $\omega_1 = 2\gamma_s B_0$, $\omega_2 = D + \gamma_s B_0$, and $\omega_3 = D - \gamma_s B_0$. In this approximation 
we have assumed that $\lambda_{0x'}^2+\lambda_{0y'}^2 \approx \lambda_{0x}^2+\lambda_{0y}^2$ (see Eqs.~\eqref{A1}-\eqref{A3}). 
At low magnetic fields, $\gamma_s B_0 \ll D$, we recover the previous result given in Eq.~\eqref{T1LowB}. Figure.~\ref{fig:Figure4}b shows the expected longitudinal relaxation rate at low temperatures for magnetic fields ranging from 0 to 1500 G. As the magnetic field increases, the longitudinal relaxation rate increases as well.

\subsection{High-temperature limit}
\begin{figure}[t!]
\includegraphics[width= 1 \linewidth]{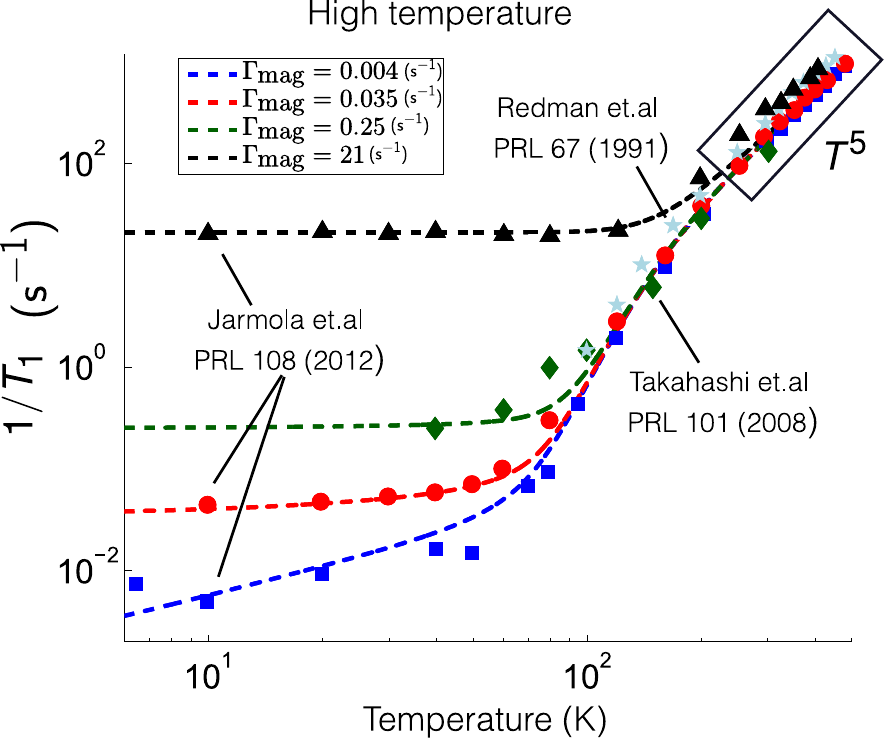}
\caption{(a) The symbols represent experimental spin relaxation rates measured for different NV$^{-}$ samples in the temperature regime 4-475 K \cite{Redman1991,Takahashi2008,Jarmola2012}. The dotted lines are the theoretical fit of the longitudinal spin relaxation rate $1/T_1$ given in Eq.~\eqref{LongitudinalRelaxationTime} for different values of the magnetic noise $\Gamma_{\mbox{\scriptsize mag}}$. The temperature at which the contribution from quasi-localized phonons and second-order phonon processes dominates is sample dependence.}
\label{fig:Figure5}
\end{figure}

In this section, we consider higher temperatures for which the relaxation rate is dominated by quasi-localized phonons
and two-phonon processes, usually for temperatures higher than 100 K. By solving the quantum master equation we
obtain that the longitudinal spin relaxation rate of $\langle S_z(t)\rangle$ is approximately given by
\begin{eqnarray}
{1 \over T_1}  &\approx &  \Gamma_{\mbox{\scriptsize mag}} + \Gamma_{\mbox{\scriptsize 1-ph}} + \Gamma_{\mbox{\scriptsize loc}} + \Gamma_{\mbox{\scriptsize 2-ph}}, \nonumber \\
&=& \Gamma_{\mbox{\scriptsize mag}} + \sum_{i=1}^{3} A_i \coth\left({\hbar \omega_i \over k_B T} \right) + {A_4 \over e^{\hbar\omega_{\mbox{\scriptsize  loc}} /k_B T}-1} \nonumber \\
&& + A_5 T^5 + A_6 T^6 + A_7  T^7.
\label{LongitudinalRelaxationTime}
\end{eqnarray}
In the above equation, $\omega_1 = 2\gamma_s B_0$, $\omega_2 = D + \gamma_s B_0$, and $\omega_3 = D - \gamma_s B_0$ are the resonant frequencies of the ground triplet states of the NV$^{-}$ center in diamond in the presence of
the static magnetic field $B_0$ along the $z$ axis, and $T$ is the temperature. Similar formulas for the longitudinal relaxation
rate were obtained phenomenologically in order to fit the experimental data for different NV$^{-}$ center samples \cite{Jarmola2012,Redman1991}. However, our work formally incorporates the phonon-induced spin relaxation rates by including the contribution of stochastic magnetic noise, direct one-phonon processes, strong interactions with quasi-localized phonon modes, and the effect of the acoustic phonons to first and second order. This is crucially different from previous works \cite{Jarmola2012,Redman1991,Takahashi2008,Astner2017}, but validates, both high and low-temperature experimental observations in which electric field fluctuations is not present (see Fig~\ref{fig:Figure5}). Our model can be useful to understand the temperature dependence of the longitudinal spin relaxation rate of other color centers in diamond. For instance, the observed $T^7$ temperature dependence of the neutral silicon-vacancy color center in diamond at high temperatures \cite{Green2017}. \par

Using experimental data from Refs. \cite{Jarmola2012,Astner2017}, we can fit our free parameter $\Gamma_{\mbox{\scriptsize mag}}$ in order to model the magnetic noise induced by magnetic impurities in samples with different
NV$^{-}$ concentrations. On the other hand, we consider that the $A_i$ parameters, which are related to the spin-phonon
coupling constants, are not sample-dependent. The $A_1$, $A_2$, and $A_3$ parameters can be found by fitting to the experimental
data at low temperature (below 1 K) \cite{Astner2017} as described in Sec.IV.A.1. The parameters $A_4$, $A_5$, $A_6$, $A_7$, and
$\omega_{\mbox{\scriptsize loc}}$ can be found by fitting to the experimental data for temperatures ranging from 4 K to 475 K \cite{Jarmola2012}. \par

Figure~\ref{fig:Figure5} shows the temperature dependence of the longitudinal relaxation rate for different samples at high temperatures.For the two-phonon processes we obtain $A_4 = 1.96 \times 10^{-3}$ s$^{-1}$, $A_5 = 2.06 \times 10^{-11}$ s$^{-1}$ K$^{-5}$, $A_6 = 9.11 \times 10^{-16}$ s$^{-1}$ K$^{-6}$, $A_7 = 2.55 \times 10^{-20}$ s$^{-1}$ K$^{-7}$, and $\omega_{\mbox{\scriptsize loc}} = 73$ meV. We observe a good agreement of our results with the experiments performed at high temperatures \cite{Jarmola2012,Redman1991,Takahashi2008}. The largest contribution at high temperatures, 300 K $< T <$ 500 K, is due to the second-order scattering (see Table I and Fig.~\ref{fig:Figure3}) usually known as the second-order Raman scattering \cite{Walker1968} which leads to the observed $1/T_1 \propto T^5$ temperature dependence \cite{Jarmola2012,Redman1991,Takahashi2008} due to the linear spin-phonon coupling to second-order. Between 50 K $<T<$ 200 K the main contribution arises from Orbach-type processes \cite{Abragam1970} which can be attributed to a strong spin-phonon interaction with a quasi-localized phonon mode with energy $\approx 73$ meV \cite{Jarmola2012}. On the other hand, the magnetic noise rate $\Gamma_{\mbox{\scriptsize mag}}$ is dominant in samples with a high NV concentration (red, green and black dashed curves in Fig.~\ref{fig:Figure5}a). Therefore, the effect of one-phonon processes (emission and absorption) can be neglected if the magnetic noise is larger than the one-phonon spin relaxation rates.  We note that we are not considering other sources of relaxation such as fluctuating electric fields, in which case a relaxation with an inverse magnetic field dependence is expected \cite{Myers2017}.

\section{Conclusions}
In summary, we have presented a microscopic model for estimating the effect of temperature on the longitudinal
relaxation rate $1/T_1$ of NV$^{-}$ centers in diamond. In this model, we introduced a general spin-phonon interaction between
the ground-state spin degree of freedom and lattice vibrations. We estimated the value of the phonon-induced spin relaxation rates by applying the Fermi golden rule to first and second order. The microscopic spin-lattice relaxation dynamics was derived from the quantum master equation for the reduced spin density operator. In the relaxation dynamics, we included the effect of a phononic bath in thermal equilibrium and dilute magnetic impurities. Acoustic and quasi-localized phonons were included in the phonon
processes in order to model a more general temperature dependence of the longitudinal relaxation rate.\par

At low temperatures, we provided a set of microscopic equations in order to study the spin-lattice relaxation dynamics
induced by one-phonon processes. In this limit and considering zero magnetic fields, $B_0 = 0$, we analytically
obtained the relaxation rate $1/T_1 = \Gamma_0(1 + 3\bar{n})$ associated with $\langle S_z^2(t) \rangle$, where $\Gamma_0$ depends on microscopic constants. This relaxation rate is in agreement with recent experiments and \textit{ab initio} calculations \cite{Jarmola2012}, as well as theoretical calculations \cite{Scott1962}. In addition, for low magnetic
fields, $\gamma_s B_0 \ll D$, we obtained the relaxation rate $1/T_1 = 2\Gamma_0 (1+2\bar{n})+\Gamma_B(1+2n_B)$ associated with 
$\langle S_z(t) \rangle$, where $\Gamma_B$ scales as $B_0^3$. \par

At high temperatures, we have modeled multiple two-phonon processes where the fitted relaxation rate associated
to $\langle S_z(t)\rangle$ is in agreement with experimental observations \cite{Jarmola2012,Redman1991,Takahashi2008}. We included both linear and bi-linear lattice interactions that lead to several different temperature scaling in a spin-boson model. In particular, for NV-centers in diamond the dominant temperature scaling is $T^5$ for temperatures larger than 200 K. Moreover, our model will be useful to evaluate the contribution of second-order phonon processes that give different
temperature scaling ($T^s, T^{s+1}, T^{s+2}$) for other spin-boson systems. The power of the temperature $s = 4\nu +2d -3$ depends on the dimension of the system and the symmetry of the lattice, where $d = 3$ and $\nu = 1/2$ for the NV$^{-}$ center.

\section{Acknowledgements}
The authors acknowledge the CONICYT UC Berkeley-Chile collaboration program. A.N. acknowledges support from Conicyt fellowship and  Gastos Operacionales of Conicyt No. 21130645. E.M. acknowledges support from Conicyt-Fondecyt 1141146. H.T.D. is funded by a Fondecyt-Postdoctoral (Grant No. 3170922). D.B. acknowledges support by the DFG through the DIP program (FO 703/2-1). J.R.M. acknowledges support from Conicyt-Fondecyt 1141185, Conicyt-PIA ACT1108, and AFOSR grant FA9550-16-1-0384.
%%%%%%%%%%%%%%%%%%% APPENDIX  %%%%%%%%%%%%%%%%%%%

\appendix

\section{Fermi golden rule}
In this section we derive the analytic form of the second-order phonon-induced spin relaxation rates introduced in Sec. III. b. To second order in time-dependent perturbation theory the transition rate between an initial $\ket{i}$ and final state $\ket{f}$ is given by
\begin{equation}
\Gamma_{i \rightarrow f} = {2 \pi \over \hbar}\left|V_{fi} +  \sum_{m}{V_{fm}V_{mi} \over E_i - E_m} \right|^2\delta(E_i - E_f), 
\label{FermiGoldenRule}
\end{equation}
where $V_{ij} = \bra{i} V \ket{j}$, with $V$ being the perturbation. In Eq.~\eqref{FermiGoldenRule} the sum over $m$ denote all possible intermediate states $\ket{m}$ for which $V_{fm}V_{mi} \neq 0$. Here, $E_i$, $E_f$, and $E_m$ are the energies of the initial, final, and intermediate states, respectively. For the Stokes transition the initial and final states are given by 
$\ket{i} = \ket{m_s,n_k,n_{k'}}$ and $\ket{f} = \ket{m_s',n_k-1,n_{k'}+1}$. If we expand the phonon part of the summation for the intermediate states $\ket{n_p,n_{p'}} = \{\ket{n_k-1,n_{k'}},\ket{n_k,n_{k'}+1}\}$, we obtain 
\begin{widetext}
\begin{eqnarray}
\Gamma_{m_s,n_k,n_{k'}}^{m_s', n_k-1,n_{k'}+1}&=& {2\pi \over \hbar}n_k(n_{k'}+1)\left|\sum_{i}g_i^{m_s',m_s} \lambda_{kk',i} 
+{1 \over \hbar}\sum_{m_s''}\sum_{i,j}\lambda_{k',i}\lambda_{k,j}\left( {g_i^{m_s',m_s''}g_j^{m_s'',m_s} \over \omega_k} - 
{g_{i}^{m_s'',m_s}g_j^{m_s',m_s''} \over \omega_{k'}}\right) \right|^2 \nonumber \\
&& \times \delta(\omega_{m_s',m_s}-\omega_k + \omega_{k'}),
\end{eqnarray}
\end{widetext}
where $g_i^{m_s,m_s'} = \bra{m_s} \hat{F}_i(\mathbf{S}) \ket{m_s'}$, and the summation over $i$ and $j$ is over $x, y, x', y', z$. Here, we have used the approximation $\omega_{k,k'} \gg \omega_{m_s,m_s''}$. By taking the continuous limit and using the density of phonon states given in Eq.~\eqref{DOS-d} we obtain 
\begin{widetext}
\begin{eqnarray}
a^{\mbox{\scriptsize Stokes}}_{m_s,m_s'}(x_D) &=& {2\pi D_0^2 \over \hbar^3 \omega_D^{4\nu+2d-2}}\int_{0}^{x_D} n(x)(n(x-x_{m_s',m_s})+1)x^{2\nu+d-1}(x-x_{m_s',m_s})^{2\nu+d-1}\nonumber \\ 
&& \hspace{2 cm}\times \left|\sum_{m_s''}\sum_{i,j}\lambda_{0i}\lambda_{0j}\left( {g_i^{m_s',m_s''}g_j^{m_s'',m_s} \over x} - {g_{i}^{m_s'',m_s}g_j^{m_s',m_s''}\over (x-x_{m_s',m_s})}\right) \right|^2 dx, \\
b^{\mbox{\scriptsize Stokes}}_{m_s,m_s'}(x_D) &=& {2\pi D_0^2 \over \hbar^2 \omega_D^{4\nu+2d-2}}\int_{0}^{x_D} n(x)(n(x-x_{m_s',m_s})+1)x^{2\nu+d-1}(x-x_{m_s',m_s})^{2\nu+d-1}\nonumber \\ 
&& \hspace{2 cm}\times 2 \mbox{Re}\left[\sum_{m_s''}\sum_{i}\sum_{i',j'}\lambda_{00i}\lambda_{0i'}\lambda_{0j'}\left( {g_{i'}^{m_s',m_s''}g_{j'}^{m_s'',m_s} \over x} - {g_{i'}^{m_s'',m_s}g_{j'}^{m_s',m_s''}\over (x-x_{m_s',m_s})}\right) \right] dx,\\
c^{\mbox{\scriptsize Stokes}}_{m_s,m_s'}(x_D) &=& {2\pi D_0^2 \left|\sum_{i} g_i^{m_s',m_s}\lambda_{00i} \right|^2 \over \hbar \omega_D^{4\nu+2d-2} }\int_{0}^{x_D} n(x)(n(x-x_{m_s',m_s})+1)x^{2\nu+d-1}(x-x_{m_s',m_s})^{2\nu+d-1}\,dx.
\end{eqnarray}
\end{widetext}
where $D_0 = \Omega \omega_D^2/(2\pi v_s^3)$ for a three-dimensional lattice, $\omega_D$ is the Debye frequency, $d$ is the dimension of the lattice, $\nu$ is the scaling of the spin-phonon coupling for acoustic phonons [see Eq.~\eqref{lambda(w)}]. Here, $x_D = \hbar D / (k_B T)$, $x_{m_s',m_s} = \hbar \omega_{m_s',m_s}/(k_B T)$, in which $\omega_{m_s',m_s} = \omega_{m_s'}-\omega_{m_s}$, $k_B$ is the Boltzmann constant, $\hbar$ is the Planck constant, and $T$ is the temperature. Similar formulas can be obtained for the other processes (Direct, Anti-Stokes and Spontaneous emission).

\section{Quantum master equation}
In this section we solve the quantum master equation for the ground state spin degree of freedom of the NV$^{-}$ center in diamond. By solving the quantum master equation given in Eq.~\eqref{QME},  for the spin populations $p_1 = \bra{m_s=1}\hat{\rho}\ket{m_s=1}$, $p_2 = \bra{m_s=0}\hat{\rho} \ket{m_s=0}$, and $p_3 = \bra{m_s=-1}\hat{\rho}\ket{m_s=-1}$, we obtain 
\begin{eqnarray}
\dot{p}_1&=& -\Gamma_1' p_1 + \Gamma_2' p_2 + \Gamma_3 p_3, \label{r11} \\
\dot{p}_2 &=& -\Gamma_4' p_2 + \Gamma_5' p_1 + \Gamma_6' p_3, \nonumber  \label{r22} \\
\dot{p}_3 &=& -\Gamma_7' p_3 + \Gamma_8 p_1 + \Gamma_9' p_2, \label{r33} 
\end{eqnarray}
where $\Gamma_i' = \Gamma_i + \Gamma_{\mbox{\scriptsize mag}}/2$, and the phonon-induced spin relaxation rates are given by
\begin{eqnarray}
\Gamma_1  &=& \Gamma_{\mbox{\scriptsize em}}^{1, \mbox{\scriptsize 1-ph}} + \Gamma_{\mbox{\scriptsize em}}^{1, \mbox{\scriptsize 2-ph}} 
                                 +\Gamma_{\mbox{\scriptsize em}}^{2, \mbox{\scriptsize 1-ph}} + \Gamma_{\mbox{\scriptsize em}}^{2, \mbox{\scriptsize 2-ph}},\\
\Gamma_2  &=&  \Gamma_{\mbox{\scriptsize abs}}^{2, \mbox{\scriptsize 1-ph}} + \Gamma_{\mbox{\scriptsize abs}}^{2, \mbox{\scriptsize 2-ph}},\\
\Gamma_3  &=&  \Gamma_{\mbox{\scriptsize abs}}^{1, \mbox{\scriptsize 1-ph}} + \Gamma_{\mbox{\scriptsize abs}}^{1, \mbox{\scriptsize 2-ph}},\\
\Gamma_4  &=&  \Gamma_{\mbox{\scriptsize abs}}^{2, \mbox{\scriptsize 1-ph}} + \Gamma_{\mbox{\scriptsize abs}}^{2, \mbox{\scriptsize 2-ph}} 
                                 +\Gamma_{\mbox{\scriptsize abs}}^{3, \mbox{\scriptsize 1-ph}} + \Gamma_{\mbox{\scriptsize abs}}^{3, \mbox{\scriptsize 2-ph}},\\
\Gamma_5  &=& \Gamma_{\mbox{\scriptsize em}}^{2, \mbox{\scriptsize 1-ph}} + \Gamma_{\mbox{\scriptsize em}}^{2, \mbox{\scriptsize 2-ph}},\\
\Gamma_6  &=& \Gamma_{\mbox{\scriptsize em}}^{3, \mbox{\scriptsize 1-ph}} + \Gamma_{\mbox{\scriptsize em}}^{3, \mbox{\scriptsize 2-ph}},\\
\Gamma_7  &=& \Gamma_{\mbox{\scriptsize abs}}^{1, \mbox{\scriptsize 1-ph}} + \Gamma_{\mbox{\scriptsize abs}}^{1, \mbox{\scriptsize 2-ph}} 
                                 +\Gamma_{\mbox{\scriptsize em}}^{3, \mbox{\scriptsize 1-ph}} + \Gamma_{\mbox{\scriptsize em}}^{3, \mbox{\scriptsize 2-ph}},  ,\\
\Gamma_8  &=& \Gamma_{\mbox{\scriptsize em}}^{1, \mbox{\scriptsize 1-ph}} + \Gamma_{\mbox{\scriptsize em}}^{1, \mbox{\scriptsize 2-ph}},\\
\Gamma_9  &=& \Gamma_{\mbox{\scriptsize abs}}^{3, \mbox{\scriptsize 1-ph}} + \Gamma_{\mbox{\scriptsize abs}}^{3, \mbox{\scriptsize 2-ph}}.
\end{eqnarray}
where $\Gamma_1 = \Gamma_5+ \Gamma_8$, $\Gamma_4 = \Gamma_2+ \Gamma_9$, and $\Gamma_7 = \Gamma_3+ \Gamma_6$, which implies that 
$\dot{p}_1+\dot{p}_2+\dot{p}_3 = 0$, and therefore, $\mbox{Tr}(\hat{\rho}) = 1$. The analytic solution for the populations $p_{i}(t)$ are determined by the following general solution
\begin{eqnarray}
\left(\begin{array}{c}
p_1(t) \\
p_2(t) \\
p_3(t) 
\end{array} \right) &=& \sum_{i=1}^{3} C_i \mathbf{v}_i e^{\lambda_i t},
\end{eqnarray}
where $\mathbf{v}_i$ and $\lambda_i$ are the eigenvectors and eigenvalues associated to the set of coupled linear equations of motions given by Eqs.\eqref{r11}-\eqref{r33}. The eigenvalues are given by
\begin{eqnarray}
\lambda_1 &=& -{1 \over 2}\left[\Gamma_{\mbox{\scriptsize mag}} + \Gamma_{\mbox{\scriptsize ph}}  + \sqrt{\Delta} \right],\\
\lambda_2 &=& -{1 \over 2}\left[\Gamma_{\mbox{\scriptsize mag}} + \Gamma_{\mbox{\scriptsize ph}}  - \sqrt{\Delta} \right],\\
\lambda_3 &=& 0,
\end{eqnarray}
where 
\begin{equation}
\Gamma_{\mbox{\scriptsize ph}} = \Gamma_1 + \Gamma_2 + \Gamma_7 = \sum_{i=1}^{3}\left(\Gamma_{\mbox{\scriptsize abs}}^i
+\Gamma_{\mbox{\scriptsize ems}}^i \right),
\end{equation}
is the total phonon-induced spin relaxation rate, and 
\begin{eqnarray}
\Delta &=& \Gamma_{\mbox{\scriptsize mag}}^2 +  2\Gamma_{\mbox{\scriptsize mag}}(\Gamma_9-\Gamma_8) + \Gamma_2^2 + \Gamma_3^2 +(\Gamma_1-\Gamma_6-\Gamma_9)^2 \nonumber \\
&& - 2\Gamma_2(\Gamma_7-\Gamma_5+\Gamma_8-\Gamma_9-\Gamma_{\mbox{\scriptsize mag}}), \nonumber \\
&& - 2\Gamma_3(\Gamma_5-\Gamma_6-\Gamma_8+\Gamma_9+\Gamma_{\mbox{\scriptsize mag}}).
\end{eqnarray}
If we consider the initial condition $\rho_{00}(0) = 1$ (ground state) and considering that $\langle S_z(t) \rangle \rightarrow 0$ when $t \rightarrow \infty$ , we finally obtain 
\begin{equation}
\langle S_z(t) \rangle =  e^{-\left(\Gamma_{\mbox{\scriptsize mag}} + \Gamma_{\mbox{\scriptsize ph}}\right) t } \sinh(\Delta t) \propto e^{-t/T_1}.
\end{equation}
Therefore, by assuming that $\left(2\Gamma_{\mbox{\scriptsize mag}} + \Gamma_{\mbox{\scriptsize ph}}\right)/2 > \Delta$, we can recover the longitudinal relaxation rate given in Eq.\eqref{LongitudinalRelaxationTime}.

\bibliographystyle{unsrt}

\end{document}